\documentclass[aps,showpacs,pre,twocolumn,letterpaper,superscriptaddress]{revtex4-1}

\usepackage{graphicx}
\usepackage{amsmath}
\usepackage{color}

\begin{document}

\title{Active crystals and their stability}

\date{\today}

\author{Andreas M.~Menzel}
\email[email: ]{menzel@thphy.uni-duesseldorf.de}
\affiliation{Institut f\"ur Theoretische Physik II: Weiche Materie, Heinrich-Heine-Universit\"at D\"usseldorf, Universit\"atsstra{\ss}e 1, D-40225 D\"usseldorf, Germany}
\affiliation{Department of Physics, Kyoto University, Kyoto 606-8502, Japan}

\author{Takao Ohta}
\affiliation{Department of Physics, Kyoto University, Kyoto 606-8502, Japan}
\affiliation{Department of Physics, Graduate School of Science, The University of Tokyo, Tokyo 113-0033, Japan}

\author{Hartmut L\"owen}
\affiliation{Institut f\"ur Theoretische Physik II: Weiche Materie, Heinrich-Heine-Universit\"at D\"usseldorf, Universit\"atsstra{\ss}e 1, D-40225 D\"usseldorf, Germany}

\begin{abstract}
A recently introduced ``active phase field crystal model'' describes the formation of ordered resting and traveling crystals in systems of self-propelled particles. Increasing the active drive, a resting crystal can be forced to perform collectively ordered migration as a single traveling object. We demonstrate here that these ordered migrating structures are linearly stable. In other words, during migration, the single crystalline texture together with the globally ordered collective motion is preserved even on large length scales. Furthermore, we consider self-propelled particles on a substrate that are surrounded by a thin fluid film. We find that in this case the resulting hydrodynamic interactions can destabilize the order. 
\end{abstract}


\maketitle

\section{Introduction}

Self-propelled particles convert chemical \cite{paxton2004catalytic,howse2007self,thutupalli2011swarming,theurkauff2012dynamic}, magnetic \cite{dreyfus2005microscopic,ghosh2009controlled}, or radiation energy \cite{jiang2010active,volpe2011microswimmers,buttinoni2012active} into directed motion. In this way, kinetic energy is intrinsically introduced into the corresponding system on the single particle level and length scale, in contrast to a process of macroscopic stirring or shaking from outside. Far-reaching consequences on the single particle behavior as well as on their collective properties emerge. 

Single artificially generated Janus particles can start to self-propel mostly due to phoretic effects when one of the two sides is selectively heated by laser light \cite{jiang2010active,volpe2011microswimmers,buttinoni2012active} or catalyzes a chemical reaction \cite{howse2007self,theurkauff2012dynamic}. 
In a liquid environment, such single colloidal particles obey rotational as well as translational diffusion. On time scales that are short when compared to the characteristic rotational diffusion time, self-propulsion shows up as directed motion. On significantly longer time scales, the overall diffusion coefficient is considerably enhanced \cite{howse2007self,tenhagen2011brownian,volpe2011microswimmers,buttinoni2012active}. Also the consequences of a possible additional particle deformability were studied extensively \cite{ohta2009deformable,ohkuma2010deformable,itino2011collective,tarama2011dynamics,tarama2012spinning,menzel2012soft,tarama2013dynamics,tarama2013dynamicsshear,yoshinaga2013spontaneous} as it may occur, for example, for self-propelled droplets on surfaces \cite{nagai2005mode,chen2009self,kitahata2011spontaneous,kitahata2012spontaneous,yoshinaga2012drift,nagai2013rotational} or in a bulk fluid \cite{thutupalli2011swarming}. Another example of artificial self-propelled particles are granular hoppers that transform vertical vibration energy into directed horizontal motion \cite{narayan2007long,kudrolli2008swarming,deseigne2012vibrated}. 

On the biological side, the conversion of chemical energy into directed motion allows bacteria to individually search for food \cite{berg1972chemotaxis} or to swim towards or away from illuminating light \cite{suematsu2011localized}. Likewise, chemical energy is used by bacteria, amoebae, or tissue cells to crawl on substrates \cite{rappel1999self,szabo2006phase,peruani2012collective}. Propulsion mechanisms are manifold and reach from the beating of a single flagellum \cite{elgeti2010hydrodynamics} via twisting deformations of whole filamental bacterial bodies \cite{wada2007model} to the synchronization of the motion of thousands of cilia \cite{uchida2010synchronization,golestanian2011hydrodynamic,osterman2011finding,uchida2011generic}. 

When many self-propelled particles act together, collective modes of migration emerge, like swarms that move as single entities \cite{rappel1999self,szabo2006phase,peruani2012collective}, traveling density bands \cite{chate2008collective,mishra2010fluctuations,menzel2011collective,ihle2013invasion}, or lanes of counterpropagating particles \cite{mccandlish2012spontaneous, wensink2012emergent, menzel2012soft}. Even turbulent states arise that are based on the energy conversion on the single particle length scale \cite{wensink2012meso, wensink2012emergent}. Already without adhesive interactions and in the absence of alignment mechanisms, clusters of self-propelled particles can form in high-density systems \cite{fily2012athermal,buttinoni2013dynamical,bialke2013microscopic}. 
These clusters appear when there is no interaction that aligns the migration directions of different particles so that they can mutually block their motion. For monodisperse two-dimensional systems, hexagonal packing can arise \cite{redner2013structure}. Confining walls can support a local clustering or trapping of self-propelled particles at the boundaries \cite{wensink2008aggregation,kaiser2012how,elgeti2013wall,kaiser2013capturing}. 

If the particle interaction is strong enough compared to their self-propulsion, then a formation of active crystals composed of self-propelled particles is to be expected. For this scenario, particle simulations demonstrated the emergence of crystal-like objects that collectively migrate as a single structure \cite{gregoire2003moving,ferrante2013elasticity,ferrante2013collective}. In contrast, the approach that we recently introduced \cite{menzel2013traveling} and analyze here in breadth is a complementary continuum one. 
It combines the description of crystalline materials by the phase field crystal model \cite{elder2002modeling,elder2004modeling,emmerich2012phase} with ideas from the Toner-Tu model \cite{toner1995long,toner1998flocks} to characterize active crystalline systems of self-propelling objects \cite{gregoire2003moving,bialke2012crystallization,redner2013structure,menzel2012soft,ferrante2013elasticity,ferrante2013collective,menzel2013unidirectional}. 
To our knowledge, there is no field-theoretical approach preceding Ref.~\cite{menzel2013traveling} that includes the translational ordering in active crystals of self-propelled particles. 

The phase field crystal model is a microscopic field approach introduced by Elder et al.\ to efficiently model processes in crystalline materials with particle resolution, but on diffusive time scales \cite{elder2002modeling,elder2004modeling}. In this way it can reproduce results from molecular dynamics simulations \cite{tupper2008phase}, yet in a computationally much more efficient way. Motivated by dynamical density functional theory \cite{elder2007phase,teeffelen2009derivation,tegze2009diffusion,jaatinen2009thermodynamics}, the phase field crystal model can in principle be viewed as a microscopic theory describing a manifold of solidification and crystallization phenomena. 
Examples are defect motion in strained crystals and multidomain structures \cite{stefanovic2006phase,stefanovic2009phase,chan2010plasticity}, epitaxial growth \cite{elder2002modeling,elder2004modeling}, crystal pinning \cite{ramos2010dynamical}, crystal growth of binary alloys \cite{elder2002modeling,elder2007phase}, branching in crystal growth \cite{tegze2011faceting}, or even crack propagation \cite{elder2004modeling}. 

Quite contrarily, the Toner-Tu model is a macroscopic hydrodynamic-like continuum description of systems of self-propelling objects \cite{toner1995long,toner1998flocks}. Local spatial -- i.e.\ translational -- order of the single particles is not resolved or explicitly taken into account. The focus in this model is on the orientational order of the migration directions and thus on the macroscopic emergence of collective motion. It was argued that, if locally the migration directions spontaneously order, long-ranged orientational order of the migration directions and therefore global collective motion into one single collective direction emerges even in two spatial dimensions \cite{toner1995long,toner1998flocks,toner2005hydrodynamics}.

Our present work starts from an ``active phase field crystal model'' that we recently introduced in a previous study \cite{menzel2013traveling}. 
It combines ideas from the two outlined theories, the equilibrium phase field crystal model and the Toner-Tu approach namely. 
We extend our previous investigation by providing for example more details about inverse active lattice structures and the migration speeds of the collectively migrating textures. 
Our main focus here is on the stability of traveling active crystals. 
We therefore perform a linear stability analysis of these structures and demonstrate linear stability even on large length scales that are not accessible by direct numerical simulations. Furthermore, for particles self-propelling on a substrate and surrounded by a thin fluid film, we include the influence of hydrodynamic interactions: they can disturb or destroy the crystalline and orientational order.

\section{Active phase field crystal model}

As a first step we derive in this section our active phase field crystal model on the basis of dynamical density functional theory for Brownian systems \cite{marconi1999dynamic,archer2004dynamical}. We consider a system of self-propelled particles in two spatial dimensions. Examples would be amoebae or tissue cells crawling on a substrate \cite{rappel1999self,szabo2006phase,peruani2012collective} or granular hoppers that move horizontally on a vibrating plate and cannot jump over each other \cite{narayan2007long,kudrolli2008swarming,deseigne2012vibrated}. Such systems can directly exchange momentum with the ground. The same should also be true for low Reynolds number swimmers in a thin film of fluid on a substrate \cite{volpe2011microswimmers,buttinoni2012active}. 

We assume that each particle migrates with an effective active drive of magnitude $v$ into an individual direction that is characterized by a unit vector $\mathbf{\hat{u}}$. The one-particle density distribution function is therefore given by $\rho(\mathbf{x},\mathbf{\hat{u}},\tau)$, with $\mathbf{x}$ the spatial coordinate and $\tau$ the time. In two spatial dimensions, the orientation of $\mathbf{\hat{u}}$ can be parameterized by a single angle $\vartheta$. Thus the dynamical equation for the one-particle density follows as \cite{wensink2008aggregation}
\begin{eqnarray}
\frac{\partial\rho(\mathbf{x},\mathbf{\hat{u}},\tau)}{\partial \tau} & = & 
  \nabla\cdot\mathbf{\tilde{D}_T}\cdot\Big( \beta\, \rho(\mathbf{x},\mathbf{\hat{u}},\tau) \,\nabla
  \frac{\delta\tilde{\mathcal{F}}}{\delta\rho(\mathbf{x},\mathbf{\hat{u}},\tau)}\Big) \nonumber\\
&&{}+\tilde{D}_r\,\partial_{\varphi}\,\Big( \beta\, \rho(\mathbf{x},\mathbf{\hat{u}},\tau)\,
  \partial_{\varphi}
  \frac{\delta\tilde{\mathcal{F}}}{\delta\rho(\mathbf{x},\mathbf{\hat{u}},\tau)}\Big) \nonumber\\
&&{}-\nabla\cdot\mathbf{\tilde{D}_T}\cdot\Big( \rho(\mathbf{x},\mathbf{\hat{u}},\tau)\,
  v\mathbf{\hat{u}}\,/\tilde{D}_{\|}\Big).
\label{eqrho}
\end{eqnarray}
In this equation $\beta=(k_BT)^{-1}$ sets the inverse thermal energy, $\mathbf{\tilde{D}_T}=\tilde{D}_{\|}\mathbf{\hat{u}}\mathbf{\hat{u}}+\tilde{D}_{\bot}(\mathbf{I}-\mathbf{\hat{u}}\mathbf{\hat{u}})$ is the translational diffusion tensor, $\tilde{D}_r$ denotes the rotational diffusion constant, and $\tilde{\mathcal{F}}$ is an equilibrium energy functional that will be specified below. The effect of self-propulsion enters through the last term on the right-hand side via the magnitude of the active drive $v$. 

Our next step is to derive dynamical equations for the particle density and for the polar orientational order of the self-propulsion directions. 
We denote the average value of the one-particle density distribution function $\rho(\mathbf{x},\mathbf{\hat{u}},\tau)$ as $\bar{\rho}$, i.e.\ 
\begin{equation}
\bar{\rho} = \frac{1}{V\Omega}\int\rho(\mathbf{x},\mathbf{\hat{u}},\tau)\,d\mathbf{\hat{u}}\,d\mathbf{x},
\end{equation}
which is a conserved quantity in a closed system. Here, $V$ is the spatial volume of the system and $\Omega$ is the surface area of the unit sphere. The modulation of the particle density $\bar{\rho}\tilde{\phi}_1(\mathbf{x},\tau)$ around $\bar{\rho}$ is obtained by integrating out the orientational degrees of freedom from $\rho(\mathbf{x},\mathbf{\hat{u}},\tau)$ and subtracting $\bar{\rho}$, 
\begin{equation}
\bar{\rho}\tilde{\phi}_1(\mathbf{x},\tau) = \frac{1}{\Omega}\int\rho(\mathbf{x},\mathbf{\hat{u}},\tau)\,d\mathbf{\hat{u}}-\bar{\rho}.
\end{equation}
Furthermore, we take into account the first orientational moment of $\rho(\mathbf{x},\mathbf{\hat{u}},\tau)$, which we denote as 
\begin{equation}
\bar{\rho}\tilde{\mathbf{P}}(\mathbf{x},\tau) = \frac{d}{\Omega}\int\rho(\mathbf{x},\mathbf{\hat{u}},\tau)\,\mathbf{\hat{u}}\,d\mathbf{\hat{u}},
\end{equation}
with $d\in\{2,3\}$ the dimensionality of the system. 
From this definition it becomes evident that the polarization field $\tilde{\mathbf{P}}(\mathbf{x},\tau)$ describes the local orientational order of the active driving directions $\mathbf{\hat{u}}$ of the self-propelled particles. 
Similarly to the procedure in Refs.~\cite{lowen2010phase,wittkowski2010derivation,wittkowski2011polar}, we expand the one-particle density with respect to its orientational dependence as 
\begin{equation} \label{rhoexpand}
\rho(\mathbf{x},\mathbf{\hat{u}},\tau)\approx\bar{\rho}+\bar{\rho}\,\tilde{\phi}_1(\mathbf{x},\tau)+\bar{\rho}\,\mathbf{\hat{u}}\cdot\tilde{\mathbf{P}}(\mathbf{x},\tau) +\dots\;. 
\end{equation} 

To proceed we insert this expansion into Eq.~(\ref{eqrho}). From the resulting equation, the different Fourier modes in terms of the angle $\varphi$ are calculated. In this way we obtain dynamical equations for $\bar{\rho}\tilde{\phi}_1$ and $\bar{\rho}\tilde{\mathbf{P}}$ similar to the ones in Refs.~\cite{lowen2010phase,wittkowski2010derivation,wittkowski2011polar}, now, however, including the active drive. Applying the approximation of constant mobility and assuming an isotropic translational diffusion $\tilde{D}_{\|}\approx \tilde{D}_{\bot}\approx \tilde{D}$, the resulting expressions are considerably simplified. We finally find 
\begin{eqnarray}
\partial_{\tau}(\bar{\rho}\tilde{\phi}_1) &=& \frac{\beta \tilde{D}\bar{\rho}}{2\pi}\nabla^2\frac{\delta\tilde{\mathcal{F}}}{\delta(\bar{\rho}\tilde{\phi}_1)} -\frac{v}{2}\,\nabla\cdot(\bar{\rho}\tilde{\mathbf{P}}), 
\label{eqpsi1dft}\\[.1cm]
\partial_{\tau}(\bar{\rho}\tilde{\mathbf{P}}) &=& \frac{\beta \tilde{D}\bar{\rho}}{\pi}\nabla^2\frac{\delta\tilde{\mathcal{F}}}{\delta(\bar{\rho}\tilde{\mathbf{P}})} -\frac{\beta \tilde{D}_r \bar{\rho}}{\pi}\frac{\delta\tilde{\mathcal{F}}}{\delta(\bar{\rho}\tilde{\mathbf{P}})} \nonumber\\[.05cm]
& & {} -v\nabla(\bar{\rho}\tilde{\phi}_1). 
\label{eqPdft}
\end{eqnarray}

In the following we specify the free energy functional $\tilde{\mathcal{F}}$. More precisely, we introduce 
\begin{equation}\label{Fsum}
\tilde{\mathcal{F}}=\tilde{\mathcal{F}}_{pfc}+\tilde{\mathcal{F}}_{\tilde{\mathbf{P}}}
\end{equation} 
as a sum of a translational part $\tilde{\mathcal{F}}_{pfc}$ and an orientational part $\tilde{\mathcal{F}}_{\tilde{\mathbf{P}}}$. 

On the one hand, 
\begin{equation}\label{Fpfc_phi}
\tilde{\mathcal{F}}_{pfc} = \int d^2\!x\,\left\{\frac{1}{2}\phi\left[ a\Delta T+\lambda(q_0^2+\nabla^2)^2\right]\phi + \frac{u}{4}\phi^4 \right\},
\end{equation}
is the free energy introduced as the phase field crystal model \cite{elder2002modeling,elder2004modeling}. Here, the order parameter $\phi$ is related to the field $\tilde{\phi}_1$ in Eq.~(\ref{rhoexpand}) via $\phi=\bar{\phi}+\phi_1$ and $\phi_1=\bar{\rho}\tilde{\phi}_1$. The parameters $a$, $\lambda$, and $u$ determine the magnitude of the energetic contributions, whereas $\Delta T$ is a measure for the temperature. Minimizing this functional for an equilibrium system leads to fluid-like states at higher temperatures or high values of $|\bar{\phi}|$, hexagonally crystalline states at lower temperatures and intermediate values of $|\bar{\phi}|$, and lamellar states at lower temperatures and low values of $|\bar{\phi}|$ \cite{elder2004modeling}. The characteristic length scale of the hexagonally crystalline and lamellar structures is determined by the magnitude of $q_0^{-1}$ in Eq.~(\ref{Fpfc_phi}). 

On the other hand, the orientational part is given by 
\begin{equation}\label{FtildeP}
\tilde{\mathcal{F}}_{\tilde{\mathbf{P}}} = \int d^2\!x\,\left\{ \frac{1}{2}\tilde{C}_1(\bar{\rho}\tilde{\mathbf{P}})^2 + \frac{1}{4}\tilde{C}_4\left[(\bar{\rho}\tilde{\mathbf{P}})^2\right]^2 \right\}.
\end{equation}
The concept behind this free energy is similar to the original approach by Toner and Tu \cite{toner1995long,toner1998flocks}. We have $\tilde{C}_4\geq0$ due to thermodynamic stability. If $\tilde{C}_1<0$ and $\tilde{C}_4>0$, the system can reduce its free energy by an orientational ordering of the self-propulsion directions, given by a nonzero solution of $\bar{\rho}\tilde{\mathbf{P}}$. This corresponds to an intrinsic tendency towards ordered collective motion. If $\tilde{C}_1>0$, orientational ordering of the self-propulsion directions always costs orientational free energy. Consequently there is a tendency to avoid collectively ordered motion. This case shows a richer behavior than the other one and will mainly be considered below. 

Predominantly using the Ramakrishnan-Yussouff expansion \cite{ramakrishnan1979first}, the phase field crystal model could be connected to density functional theory \cite{elder2007phase,teeffelen2009derivation}; for a review see Ref.~\cite{emmerich2012phase}. Expressions for the phenomenological parameters in terms of microscopic correlation functions could be derived. 
Afterwards the procedure was extended to the case of an additional orientational order \cite{lowen2010phase,wittkowski2010derivation,wittkowski2011polar,wittkowski2011microscopic}. Likewise microscopic expressions for the corresponding phenomenological parameters were listed in these references, and we do not repeat them here. 

To finally obtain our dynamical equations, we introduce Eqs.~(\ref{Fsum})--(\ref{FtildeP}) into Eqs.~(\ref{eqpsi1dft}) and (\ref{eqPdft}). Moreover we apply the rescaling rules \cite{elder2004modeling} 
\begin{equation}
\phi=\left(\lambda q_0^4 u^{-1}\right)^{\frac{1}{2}}\psi, \quad a\Delta T=\lambda q_0^4\,\varepsilon, \quad \mathbf{x} = q_0^{-1}\mathbf{r}. 
\end{equation}
This rescales $\phi=\bar{\phi}+\phi_1$ to $\psi=\bar{\psi}+\psi_1$. All lengths from now on are measured in units of the characteristic length scale $q_0^{-1}$ of the phase field crystal structures. 

An additional simplification of the notation follows from the further rescaling 
\begin{eqnarray}
\tau & = & \frac{2\pi}{\beta\tilde{D}\bar{\rho}}\frac{1}{\lambda q_0^6}\,t, \\
v & = & \frac{\beta\tilde{D}\bar{\rho}}{2^{\frac{1}{2}}\pi}\,\lambda q_0^5 \,v_0, \\
\tilde{\mathbf{P}} & = & \frac{1}{\bar{\rho}}\, q_0^2 \left(2\frac{\lambda}{u}\right)^{\!\frac{1}{2}}\,\mathbf{P}, \\
\tilde{C}_1 & = & \frac{\lambda q_0^4}{2}\,C_1, \\[0.05cm]
\tilde{C}_4 & = & \frac{u}{4}\,C_4, \\[.1cm]
\tilde{D}_r & = & \tilde{D}q_0^2\,D_r.
\end{eqnarray}
At the end of this procedure we obtain our dynamical order parameter equations in the form \cite{menzel2013traveling}
\begin{eqnarray}
\partial_{t}\psi_1 &=& \nabla^2\Big\{\!\left[\varepsilon+(1+\nabla^2)^2+3\bar{\psi}^2\right]\psi_1 
+ 3\bar{\psi}\psi_1^2+\psi_1^3\,\Big\} \nonumber\\
& & {} - v_0\,\nabla\cdot\mathbf{P}, \label{eqpsi1}\\[.1cm]
\partial_{t}\mathbf{P} &=& \nabla^2\left( C_1\mathbf{P}+C_4\mathbf{P}^2\mathbf{P} \right) - D_r\left( C_1\mathbf{P}+C_4\mathbf{P}^2\mathbf{P} \right)\nonumber\\[.05cm]
& & {} - v_0\nabla\psi_1.
\label{eqP}
\end{eqnarray}
The field $\psi_1$ is a measure for the local particle density, whereas the field $\mathbf{P}$ measures the local orientational order of the active driving directions of the self-propelled particles. $\varepsilon$ controls the temperature, $\bar{\psi}$ the mean density, and $v_0$ the strength of the active drive of the individual particles. The ordering behavior of the active driving directions of the self-propelled particles is determined by $C_1$ and $C_4$, whereas $D_r$ describes the orientational diffusion of the active drive directions.

\section{Predictions of the model}

\begin{figure*}
\centerline{\includegraphics[width=17.cm]{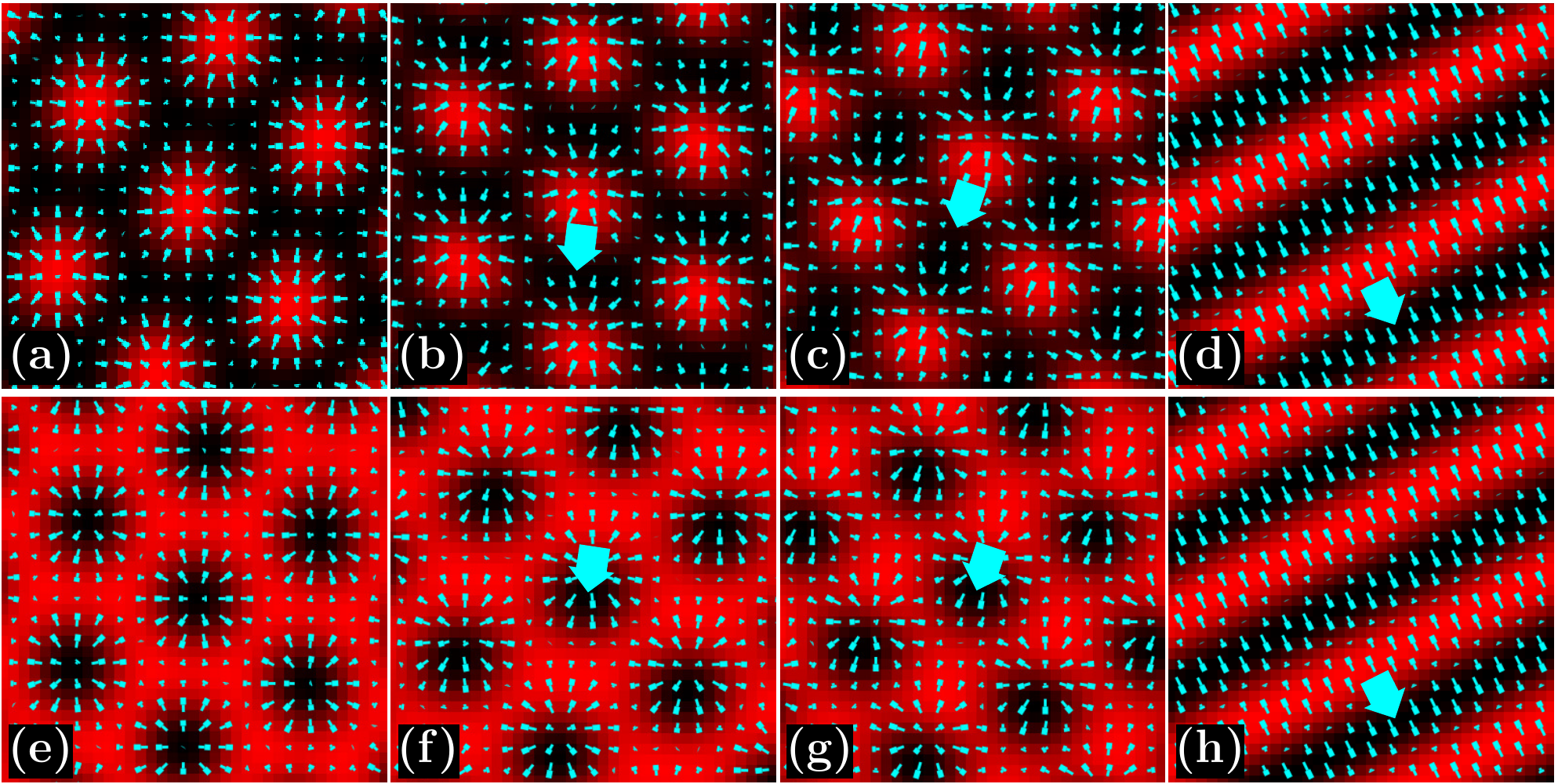}}
\caption{Snapshots of the order parameter fields illustrating the different phases that appear at $(\varepsilon,C_1,C_4,D_r)=(-0.98,0.2,0,0.5)$ for increasing active drive $v_0$. The mean density is $\bar{\psi}=-0.4$ in the upper row (a)--(d), and $\bar{\psi}=0.4$ in the lower row (e)--(h). From left to right the active drive $v_0$ of the individual particles increases from $v_0=0.1$~(a,e) via $v_0=0.5$~(b,f) and $v_0=1$~(c,g) to $v_0=1.9$~(d,h). In all snapshots, brighter color corresponds to higher densities $\psi_1$, whereas the thin bright needles illustrate the polarization field $\mathbf{P}$ pointing from the thick to the thin ends. Starting from a resting hexagonal structure in panel (a), increasing the active drive first leads to a traveling hexagonal texture (b), then a traveling quadratic structure (c), and finally traveling lamellae (d). When starting from a resting honeycomb texture (e), increasing the active drive first forces the honeycomb lattice to travel (f), then deforms the structure into an inverted squared texture (g), and finally leads to travelling lamellae again (h). The thick bright arrows indicate the predominant direction of migration of the structures in panels (b)--(d) and (f)--(h). Only a fraction of the numerical calculation box is shown.}
\label{snapshots}
\end{figure*}

As noted above, the equilibrium phase field crystal model based on Eq.~(\ref{Fpfc_phi}) shows hexagonally crystalline structures in a certain range of temperature $\varepsilon$ and mean density $\bar{\psi}$. Changing the temperature or mean density, hexagonal textures can be transformed into lamellar ones. In the non-equilibrium case considered here, 
such a transition can also be achieved by increasing the active drive $v_0$ of the individual self-propelled particles \cite{menzel2013traveling}. This scenario is depicted in the upper row of Fig.~\ref{snapshots}. 

The snapshots shown in the figure were obtained by numerically solving Eqs.~(\ref{eqpsi1}) and (\ref{eqP}), starting from random initial conditions. 
We start with a low value of the active drive $v_0$ and a hexagonal structure in panel (a). For the parameter values listed in the figure caption, density peaks form that are symmetrically located within ``+1''-defects of the polarization field. The overall structure remains at rest. This changes with increasing $v_0$, when the symmetry of the density peak positions with respect to the ``+1''-defects of the polarization field is spontaneously broken. This spontaneous shift of the density peaks out of the centers of the ``+1''-defects of the polarization field represents the symmetry breaking necessary for a net active propulsion to emerge. As a whole, the structure starts to migrate (travel) collectively~(b). Further increasing $v_0$ deforms the traveling hexagonal crystal to a traveling rhombic and then to a traveling quadratic crystal~(c). Finally it leads to a transition to traveling lamellae~(d). 

Typically, when thinking of crystalline textures in this way, we would identify each density peak with one particle. However, also the situation of cluster crystals is included by our approach. In that case several particles occupy one lattice site \cite{likos1998freezing}. Such active cluster crystalline structures were observed before in a system of deformable self-propelled particles \cite{menzel2012soft} and in a modified Vicsek model confined between two parallel walls \cite{menzel2013unidirectional}. Furthermore, traveling crystalline textures have been observed previously in three-component reaction-diffusion systems \cite{okuzono2001self,sugiura2002time,okuzono2003traveling}. 

When we thus identify each density peak in the top row of Fig.~\ref{snapshots} with a cloud of self-propelled particles, we can similarly interpret the corresponding active inverted textures in the bottom row. In each column of Fig.~\ref{snapshots} the two snapshots can be mapped onto each other by noting a special symmetry relation of Eqs.~(\ref{eqpsi1}) and (\ref{eqP}). The equations remain invariant under the simultaneous transformations $\bar{\psi}\rightarrow -\bar{\psi}$, $\psi_1\rightarrow -\psi_1$, and $\mathbf{P}\rightarrow -\mathbf{P}$. Consequently the series of patterns in the bottom row of Fig.~\ref{snapshots} follows directly as an inversion of the top row patterns. We will concentrate on the characterization of hexagonal, rhombic, and quadratic crystalline textures in the following. Nevertheless the results apply equally for the inverted structures due to the symmetry property of the dynamical equations. Active honeycomb textures are formed for example by flagellated marine bacteria \cite{thar2002conspicuous,thar2005complex}. 

To quantify the collective motion of the active crystals, we tracked the migration of each individual density peak in the sample. For this purpose, the center of each density peak was determined at fixed time intervals. On the one hand, this time interval was short enough so that identical density peaks could be identified before and after each migration step. On the other hand, it should be long enough so that the migration step is larger than the mesh size of the calculation grid. It was shown previously that -- at least for the finite system sizes investigated -- a collectively traveling single crystal develops from the random initial conditions via an intermediate multidomain structure \cite{menzel2013traveling}. At the end of this coarsening process towards the active single crystal, all density peaks migrate collectively into the same direction. 

\begin{figure}
\centerline{\includegraphics[width=8.5cm]{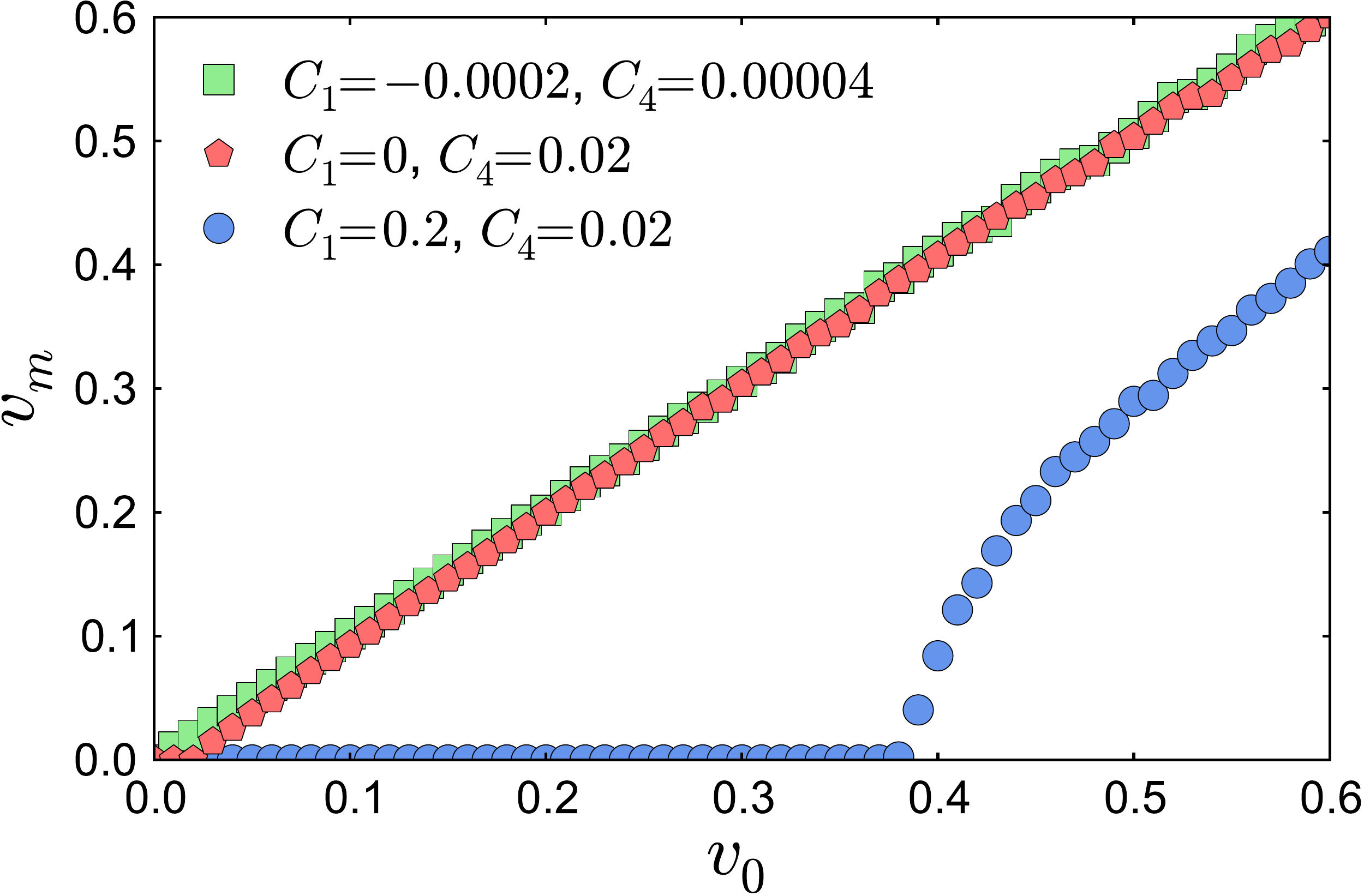}}
\caption{Collective migration speed $v_m$ of the single crystals as a function of the active drive $v_0$ of the individual particles at $(\varepsilon,\bar{\psi},D_r)=(-0.98,-0.4,0.5)$. For $C_1<0$ and $C_4>0$ (squares) the directions of active drive of the individual particles order spontaneously and the structures travel collectively with $v_m=v_0$. A nonzero threshold appears for $C_1=0$ and $C_4>0$ (rhombi). At $C_1>0$ and $C_4\geq0$ (circles) the directions of active drive of the individual particles do not order spontaneously. Then the structures are at rest ($v_m=0$) below a finite threshold value $v_{0,c}$; they start to travel collectively with nonzero $v_m<v_0$ above the threshold.}
\label{vmvsv0}
\end{figure}

The migration speed $v_m$ of the resulting active single crystal follows from a sample average of the peak velocity magnitudes of all density peaks, $v_m=\sum_{i=1}^{N_p} \|\mathbf{v}_i\|/{N_p}$. Here $N_p$ denotes the total number of all density peaks in the sample. The velocities $\mathbf{v}_i$ of the individual density peaks, with $i=1,..,N_p$, follow from the tracking procedure described above. Corresponding results are depicted in Fig.~\ref{vmvsv0} for characteristic values of the parameters $C_1$ and $C_4$. As mentioned above, these parameters control the spontaneous orientational ordering of the migration directions of the individual self-propelled particles. 
We plot in Fig.~\ref{vmvsv0} the collective migration speed $v_m$ of the single crystal as a function of the active drive $v_0$ of the individual self-propelled particles. 

For $C_1>0$ the active driving directions of the individual particles do not order spontaneously. At low values of the active drive $v_0$, the overall hexagonally crystalline structure remains at rest, i.e.\ $v_m=0$. This corresponds to the situation in Fig.~\ref{snapshots}(a). Only at a nonzero threshold value $v_{0,c}$ do the crystals start to travel collectively, see also Fig.~\ref{snapshots}(b). Here, the nonzero value of $C_4>0$ only plays a minor quantitative role, and a similar curve is obtained for $C_4=0$. 

In contrast, there is no threshold when the active driving directions of the individual particles order spontaneously, i.e.\ $C_1<0$ and $C_4>0$. The whole crystalline structure migrates at each nonzero value of the active drive $v_0$ of the individual particles. Even more, the whole crystal propels collectively at the speed of the individual particles, i.e.\ $v_m=v_0$. In addition to this curve of negative $C_1$, we show in Fig.~\ref{vmvsv0} an intermediate case of $C_1=0$ and $C_4>0$. There, the nonlinear $C_4$-term leads to a small residual threshold behavior at low values of $v_0$. 

During the remaining part of this paper we will be concerned with the case of $C_1>0$ and $C_4=0$. It shows the richer behavior due to the emergence of resting as well as traveling structures that are separated by the finite threshold value $v_{0,c}$.

\section{Linear stability analysis}

Introducing a hydrodynamic-like continuum theory, Toner and Tu studied the collective motion of flocks of self-propelling objects \cite{toner1995long,toner1998flocks,toner2005hydrodynamics}. Their focus was mainly on the orientational ordering of the self-propulsion directions, and not on a translational, possibly crystalline order of the individual objects. The theory contains nonlinear convective terms due to self-propulsion. Analyzing their model, Toner and Tu found that long-ranged orientational order of the self-propulsion directions can arise in two spatial dimensions \cite{toner1995long,toner1998flocks,toner2005hydrodynamics}, in contrast to what is known for orientational ordering in equilibrium systems \cite{mermin1966absence}. 

Such nonlinear convective terms that could lead to long-ranged orientational order of the self-propulsion directions are absent in our approach. Therefore the question arises whether a very large crystal will still migrate as a single structure into one direction, or whether it will break up into domains of different migration directions. 

Instead of the nonlinear convective terms, however, our approach contains a kind of elastic interaction. The phase field crystal free energy functional $\mathcal{F}_{pfc}$ of Eq.~(\ref{Fpfc_phi}) acts in the form of a chemical potential in the dynamical equation (\ref{eqpsi1}) for the density. It enforces a locally spatially periodic arrangement of the density peaks with a characteristic wave number. Deviations from this order are energetically penalized. 

It has been demonstrated in particle simulations \cite{ferrante2013elasticity,ferrante2013collective} that solely elastic interactions between self-propelled particles, without any explicit alignment interaction \cite{vicsek1995novel}, can induce global collective motion into one common migration direction. There the elastic interactions guided the particles to find a global mode of migration, even under the influence of moderate stochastic noise. In this way, elasticity enforced global collective motion. Due to the finiteness of the system sizes it is difficult, however, to make a rigid statement about what happens on very large length scales on the basis of the particle simulations. 

Our continuum field approach offers an advantage in this respect. The explicit dynamical equations (\ref{eqpsi1}) and (\ref{eqP}) contain all length scales. In principle, their investigation is not limited to a certain system size. An arbitrary large system can be analyzed. In reality, however, to get an explicit solution, we have to solve the equations on a finite numerical lattice. Nevertheless, we can extract from our equations the information of whether an arbitrarily large single crystal with a single migration direction is linearly stable. Our procedure will be described in the following. 
As mentioned before, we confine ourselves for this purpose to the case in which the self-propulsion directions do not spontaneously order, i.e.\ $C_1>0$ ($C_4=0$). 

As a first step we linearize the dynamical equations (\ref{eqpsi1}) and (\ref{eqP}) with respect to a small perturbation $\delta \psi_1(\mathbf{r},t)$ and $\delta \mathbf{P}(\mathbf{r},t)$ of the order parameter fields, 
\begin{eqnarray}
\psi_1(\mathbf{r},t) & = & \psi_{10}(\mathbf{r},t) + \delta \psi_1(\mathbf{r},t), \\
\mathbf{P}(\mathbf{r},t) & = & \mathbf{P}_{\!0}(\mathbf{r},t) + \delta \mathbf{P}(\mathbf{r},t). 
\end{eqnarray}
Here, $\psi_{10}(\mathbf{r},t)$ and $\mathbf{P}_{\!0}(\mathbf{r},t)$ correspond to the solutions obtained from directly solving numerically the dynamical equations (\ref{eqpsi1}) and (\ref{eqP}). These solutions describe, on the finite size of the numerical calculation grid, a single crystalline structure as illustrated for example in Fig.~\ref{snapshots}. 
In our case, the converged steady-state solutions $\psi_{10}(\mathbf{r},t)$ and $\mathbf{P}_{\!0}(\mathbf{r},t)$, around which we perturb the system, are spatially inhomogeneous. 
This significantly complicates the analysis as will be shown below. 

From the linearization we obtain
\begin{eqnarray}
\partial_{t}\delta\psi_1 &=& \nabla^2\Big\{\!\left[\varepsilon+(1+\nabla^2)^2+3\bar{\psi}^2\right]\delta\psi_1 
+ 6\bar{\psi}\psi_{10}\delta\psi_1 \nonumber\\
& & {} \qquad+3\psi_{10}^2\delta\psi_1 \Big\}  - v_0\,\nabla\cdot\delta\mathbf{P}, \label{eqdeltapsi1}\\[.1cm]
\partial_{t}\delta\mathbf{P} &=& \nabla^2 C_1\delta\mathbf{P} - D_r C_1\delta\mathbf{P}
 - v_0\nabla\delta\psi_1.
\label{eqdeltaP}
\end{eqnarray}
Since the coefficients of the perturbations in the second equation are constant, the system of equations can be significantly simplified in Fourier space. 
We apply the transformations in the form 
\begin{eqnarray}
\delta\psi_1(\mathbf{q},\omega) & = & \int\! d^2r\, e^{-i\mathbf{q}\cdot\mathbf{r}} \int\! dt\, e^{-i\omega t}\; \delta\psi_1(\mathbf{r},t),\\
\delta\mathbf{P}(\mathbf{q},\omega) & = & \int\! d^2r\, e^{-i\mathbf{q}\cdot\mathbf{r}} \int\! dt\, e^{-i\omega t}\; \delta\mathbf{P}(\mathbf{r},t).
\end{eqnarray}
Eqs.~(\ref{eqdeltapsi1}) and (\ref{eqdeltaP}) are transformed accordingly. Then the second equation can be solved for $\delta\mathbf{P}(\mathbf{q},\omega)$, which is inserted into the first equation. In this way, the system of Eqs.~(\ref{eqdeltapsi1}) and (\ref{eqdeltaP}) is reduced to a single equation on the scalar order parameter field $\delta\psi_1(\mathbf{q},\omega)$, which reads
\begin{eqnarray}
\lefteqn{i\omega\,\delta\psi_1(\mathbf{q},\omega) =} \nonumber\\
&& {} -q^2\Big\{\, \big[\varepsilon+(1-q^2)^2+3\bar{\psi}^2\big]\,\delta\psi_1(\mathbf{q},\omega) \nonumber\\
&& {} +\frac{6\bar{\psi}}{(2\pi)^3}\int\!d^2q'\int\!d\omega'\; \psi_{10}(\mathbf{q}-\mathbf{q}',\omega-\omega')\delta\psi_1(\mathbf{q}',\omega') \nonumber\\
&& {} +\frac{3}{(2\pi)^3}\int\!d^2q'\int\!d\omega'\; \psi_{10}^2(\mathbf{q}-\mathbf{q}',\omega-\omega')\delta\psi_1(\mathbf{q}',\omega') \,\Big\} \nonumber\\
&& {} -\frac{q^2v_0^2}{C_1(D_r+q^2)+i\omega}\,\delta\psi_1(\mathbf{q},\omega).
\label{stabaneq}
\end{eqnarray}

As we can see, the analysis becomes nonlocal in Fourier space. The reason is the spatial modulation of the traveling or resting unperturbed structures, as they are for example depicted in Fig.~\ref{snapshots}. In principle, Eq.~(\ref{stabaneq}) represents an infinite system of equations in which all modes $\delta\psi_1(\mathbf{q},\omega)$ of different $\mathbf{q}$ and $\omega$ are coupled. A similar situation occurs in the study of microphase-separated diblock copolymer systems Ref.~\cite{laradji1997stability}. There, however, an equilibrium situation is investigated, which allows a different treatment of the problem. 

Fortunately the situation can be simplified by exploiting the regular nature of the unperturbed states $\psi_{10}(\mathbf{r},t)$. First, the structures form a steady state in the sense that they migrate as a single object with a constant migration velocity $\mathbf{v_m}$. In the co-moving frame, the single crystals are at rest. Second, the spatial modulation of the unperturbed structures is periodic. Using these observations, Eq.~(\ref{stabaneq}) can be split into discrete sets of equations in Fourier space, each of them still being infinite, however. A numerical evaluation is required to handle the problem. 

Since the unperturbed structures $\psi_{10}(\mathbf{r},t)$ form a steady state, we can write them in the form
\begin{equation}\label{steadystate}
\psi_{10}(\mathbf{r},t) = \psi_{10}(\mathbf{r}-\mathbf{v_m}t).
\end{equation}
This relation translates itself into Fourier space in the following way: 
\begin{eqnarray}
\label{psi10co}
\psi_{10}(\mathbf{q},\omega) &=& 2\pi\delta(\omega+\mathbf{q}\cdot\mathbf{v_m})\psi_{10}(\mathbf{q}), \\
\label{psi10sqco}
\psi_{10}^2(\mathbf{q},\omega) &=& 2\pi\delta(\omega+\mathbf{q}\cdot\mathbf{v_m})\psi_{10}^2(\mathbf{q}). 
\end{eqnarray}
Introducing these expressions into Eq.~(\ref{stabaneq}) decouples the modes in the frequencies $\omega$, which significantly simplifies the problem. 

For the second simplification, we use the fact that the unperturbed structures are spatially periodic. They are resting or traveling crystals. In Fourier space they are therefore characterized by a reciprocal lattice of a discrete set of wave vectors. Translating in reciprocal space the whole structure by a reciprocal lattice vector maps the positions of the reciprocal lattice points onto themselves. In this sense, the reciprocal lattice is closed by itself under such translations. The subtractions $\mathbf{q}-\mathbf{q}'$ in Eq.~(\ref{stabaneq}) correspond to these translations. As a consequence, the set of equations (\ref{stabaneq}) splits into different decoupled discrete subsets of equations. Within one such set, all modes are connected via addition or subtraction of the reciprocal lattice vectors of the unperturbed structure. Still each set generally contains infinitely many modes, but the discretization already implies a major reduction. 

\begin{figure}
\centerline{\includegraphics[width=8.5cm]{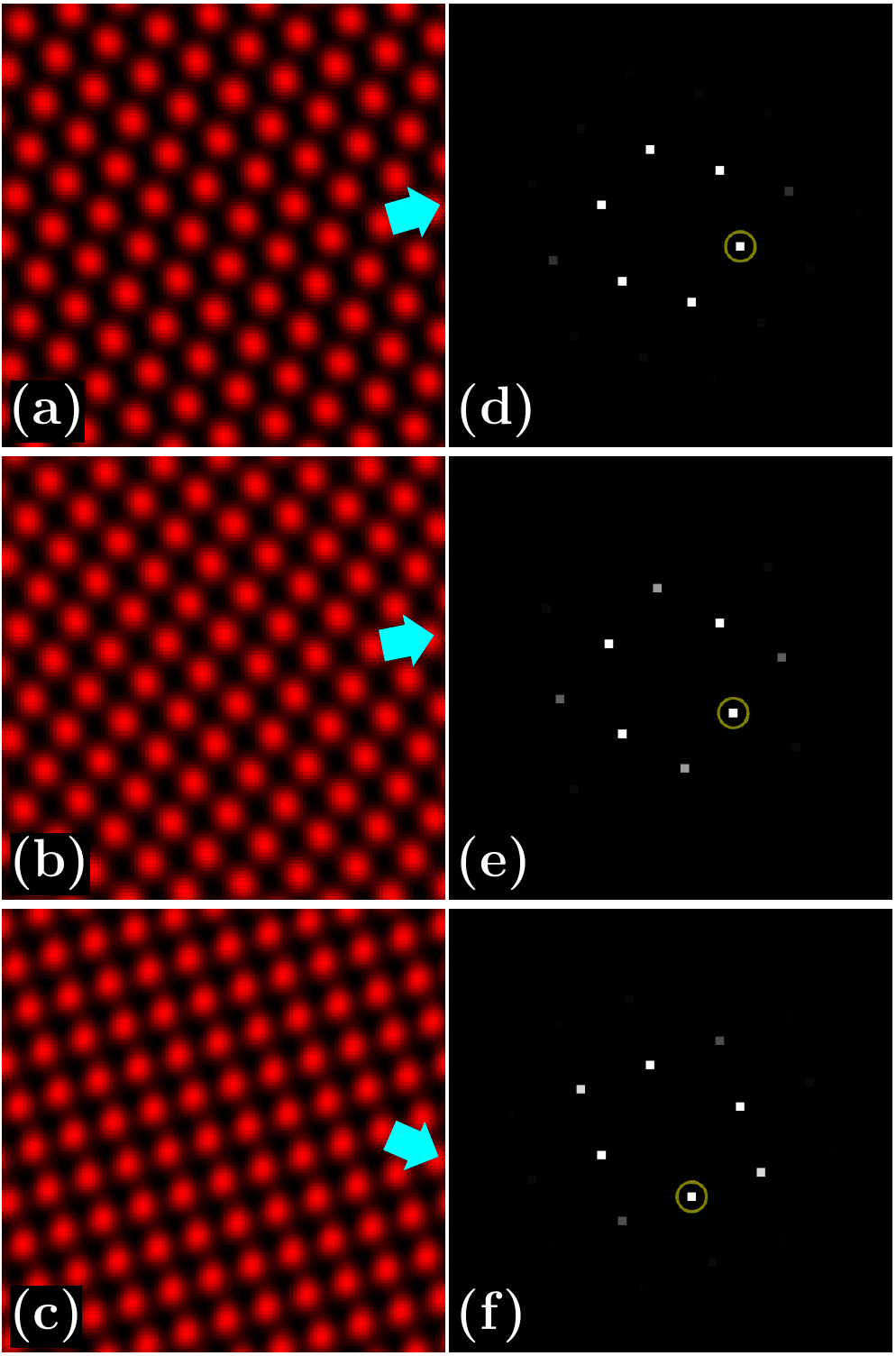}}
\caption{Left column: snapshots of crystalline structures at $(\varepsilon,\bar{\psi},C_1,C_4,D_r)=(-0.98,-0.4,0.2,0,0.5)$ and (a) $v_0=0.5$ for a traveling hexagonal crystal, (b) $v_0=0.7$ for a traveling rhombic crystal, and (c) $v_0=0.9$ for a traveling quadratic crystal. Only a fraction of the numerical calculation box is shown. Right column: corresponding power spectra in two-dimensional Fourier space. A distorted hexagonal lattice structure is obtained in Fourier space in panel (d) from the traveling hexagonal case. Panel (e) gives the power spectrum for the traveling rhombic crystal. Finally, the square pattern in real space transforms itself into a square pattern of the two-dimensional power spectrum (f). The intensity of the peaks in the power spectrum is represented by their brightness and was partially rescaled for better visualization. Circles mark the directions of the reciprocal lattice vectors of smallest magnitude $\mathbf{G}_1$. The examples of linear stability analysis against linear dynamical instabilities that are shown in Fig.~\ref{logdeterminantdynamic} were performed in these directions.}
\label{FTstructure}
\end{figure}

In the following we investigate three different structures: a traveling hexagonal crystal, a traveling rhombic crystal, and a traveling quadratic crystal. 
As mentioned above, the textures are obtained in this sequence when increasing the active drive $v_0$. 
On the left-hand side, Fig.~\ref{FTstructure} shows snapshots of the steady-state structures in real space, i.e.\ $\psi_{10}(\mathbf{r})$. On the right-hand side, the corresponding power spectra $|\psi_{10}(\mathbf{q})|$ in two-dimensional Fourier space are included. 

The different textures -- hexagonal, rhombic, and quadratic -- can be identified by eye from the real-space plots. Also the spatial arrangements of the inner intensity peaks on the right-hand side of Fig.~\ref{FTstructure} reflect these symmetries. However, when we take into account the magnitude of the intensities, we see that the collective migration breaks the symmetry. Most obviously, there is no four-fold symmetry in the intensity magnitudes of Fig.~\ref{FTstructure} (f) for the quadratic texture. Likewise, the single real-space density peaks in Fig.~\ref{FTstructure} (c) appear a little distorted. The direction of collective migration marks the orientation of the remaining symmetry axis. 

We can see that the magnitude of the peaks in the power spectrum, coded by their brightness in Fig.~\ref{FTstructure}, significantly decays with increasing wave number. As a consequence we introduce a positive threshold value $\psi_{th}$. $\psi_{10}(\mathbf{r})$ is approximated by only those modes that satisfy $|\psi_{10}(\mathbf{q})|>\psi_{th}$. The same procedure is applied to $\psi_{10}^2(\mathbf{q})$, i.e.\ the Fourier transform of $\psi_{10}^2(\mathbf{r})$. 

For each investigated structure, the above procedure must be performed only once in practice. We numerically iterate Eqs.~(\ref{eqpsi1}) and (\ref{eqP}) forward in time 
starting from random initial conditions. When the system has reached the steady state, $\psi_{1}$ is used as an input for the unperturbed states $\psi_{10}(\mathbf{r})$ and $\psi_{10}^2(\mathbf{r})$. After performing the Fourier transform, the values $\psi_{10}(\mathbf{q})$ and $\psi_{10}^2(\mathbf{q})$ that satisfy the above threshold criterion together with their locations $\mathbf{q}$ in the reciprocal space are stored in an ordered list structure for the further analysis. We also numerically measure $\mathbf{v_m}$ during this initial numerical iteration and use this value in our further evaluation. 
(It is a challenging task to calculate the collective migration velocity $\mathbf{v_m}$ analytically. A corresponding attempt in one spatial dimension that is restricted to the threshold vicinity is included in the appendix.) 

Our intention is to perform a linear stability analysis of the traveling single crystal at long wave lengths, i.e.\ at small wave numbers $\|\mathbf{q}\|$. Direct numerical solution of Eqs.~(\ref{eqpsi1}) and (\ref{eqP}) showed that the structures are stable on the length scale of the size of the numerical box. But the question of what happens on larger length scales has so far remained unanswered. In other words, we are interested in the question whether the single crystal will break up into a multidomain texture for large system sizes. 


Below, we denote the reciprocal lattice vectors of the unperturbed structure as $\mathbf{G}_i$, with $i$ a labeling index. They localize the positions of the spectral points in the right column of Fig.~\ref{FTstructure}. The reciprocal lattice vector of smallest magnitude is called $\mathbf{G}_1$. 
We probe a square region in reciprocal space bordered at two opposing edges by $-\mathbf{G}_1$ and $+\mathbf{G}_1$. This region covers the first Brillouin zone. 

All wave vectors $\mathbf{q}$ lying in the first Brillouin zone belong to a different discrete coupled subset of equations spanned by the nonlocal terms in Eq.~(\ref{stabaneq}). We cannot connect two different wave vectors in this region by adding or subtracting reciprocal lattice vectors. 
Still, these subsets of equations are infinite. We have seen, however, that the coupling strengths $|\psi_{10}(\mathbf{q}-\mathbf{q'})|$ and $|\psi_{10}^2(\mathbf{q}-\mathbf{q'})|$ that are responsible for the nonlocal coupling in Eq.~(\ref{stabaneq}) decay with increasing distance $\|\mathbf{q}-\mathbf{q'}\|$. Therefore we introduce a cutoff distance $q_{max}>0$ and only include modes that satisfy $\|\mathbf{q}-\mathbf{q'}\|<q_{max}$ in the subset of equations for each wave number $\mathbf{q}$. In this way, the discrete subsets of equations become finite and can be handled numerically. 

In practice, we sample the square region between $-\mathbf{G}_1$ and $+\mathbf{G}_1$ in small wave vector steps. At each $\mathbf{q}$ in this region, we build up the discrete set of equations described above. For this purpose, we numerically search for all modes $\delta\psi_1(\mathbf{q}',\omega')$ with wave vectors $\mathbf{q}'$ that can be obtained by iteratively adding or subtracting the reciprocal lattice vectors that predominantly contribute to $\psi_{10}$ and $\psi_{10}^2$ (those were previously stored in a list when calculating the spectra, see above). Only modes $\delta\psi_1(\mathbf{q}',\omega')$ that satisfy the condition $\|\mathbf{q}-\mathbf{q'}\|<q_{max}$ are included. 

Eq.~(\ref{stabaneq}) leads for each wave vector $\mathbf{q}$ and frequency $\omega$ to a linear system of equations for all modes $\delta\psi_1(\mathbf{q}',\omega')$ that are coupled to $\delta\psi_1(\mathbf{q},\omega)$. In practice, our cut-off values are chosen such that the coupling to several hundred modes is included for each wave vector $\mathbf{q}$. The coefficients of this system of equations are calculated and stored in matrix form. For practical reasons, to obtain a matrix of real coefficients, we consider the real and imaginary parts of the perturbations $\delta\psi_1(\mathbf{q}',\omega')=\Re\{\delta\psi_1(\mathbf{q}',\omega')\}+i\Im\{\delta\psi_1(\mathbf{q}',\omega')\}$ as independent variables. Each equation in the system of equations (\ref{stabaneq}) is split into its real and imaginary part, which doubles the number of equations. 
Finally, the determinant $\mathcal{D}(\mathbf{q},\omega)$ of the resulting matrix is calculated by standard numerical procedures \cite{press1992numerical}. 
At the end of this procedure, a vanishing determinant $\mathcal{D}(\mathbf{q},\omega)$ signals a linear instability of the single crystal with respect to the perturbation $\delta\psi_1(\mathbf{q},\omega)$. 

Before continuing with the results, we add a technical remark to explain the appearance of the plots included below. Due to the many modes that are coupled in each case, the number of multiplications during the process of obtaining the determinant $\mathcal{D}(\mathbf{q},\omega)$ is very large. In fact, $\mathcal{D}(\mathbf{q},\omega)$ often exceeds the size that can be processed on the computer. To circumvent this problem, we rescale the factors in the final multiplication by an adjusted constant positive number. Since we are only interested in the question whether $\mathcal{D}(\mathbf{q},\omega)$ vanishes (or changes sign) this is a legal procedure. Nevertheless, the absolute values of $\mathcal{D}(\mathbf{q},\omega)$ should not be interpreted any more. Furthermore, we calculate and plot the logarithm of the determinant $\ln[\mathcal{D}(\mathbf{q},\omega)]$. 

In a first step, we consider static instabilities in the co-moving frame. That is, we set $\omega=-\mathbf{q}\cdot\mathbf{v_m}$ (and likewise $\omega'=-\mathbf{q}'\cdot\mathbf{v_m}$) in Eq.~(\ref{stabaneq}). Thus $\omega$ does not describe an additional degree of freedom but is determined by the wave vector $\mathbf{q}$. 

\begin{figure}
\centerline{\includegraphics[width=8.5cm]{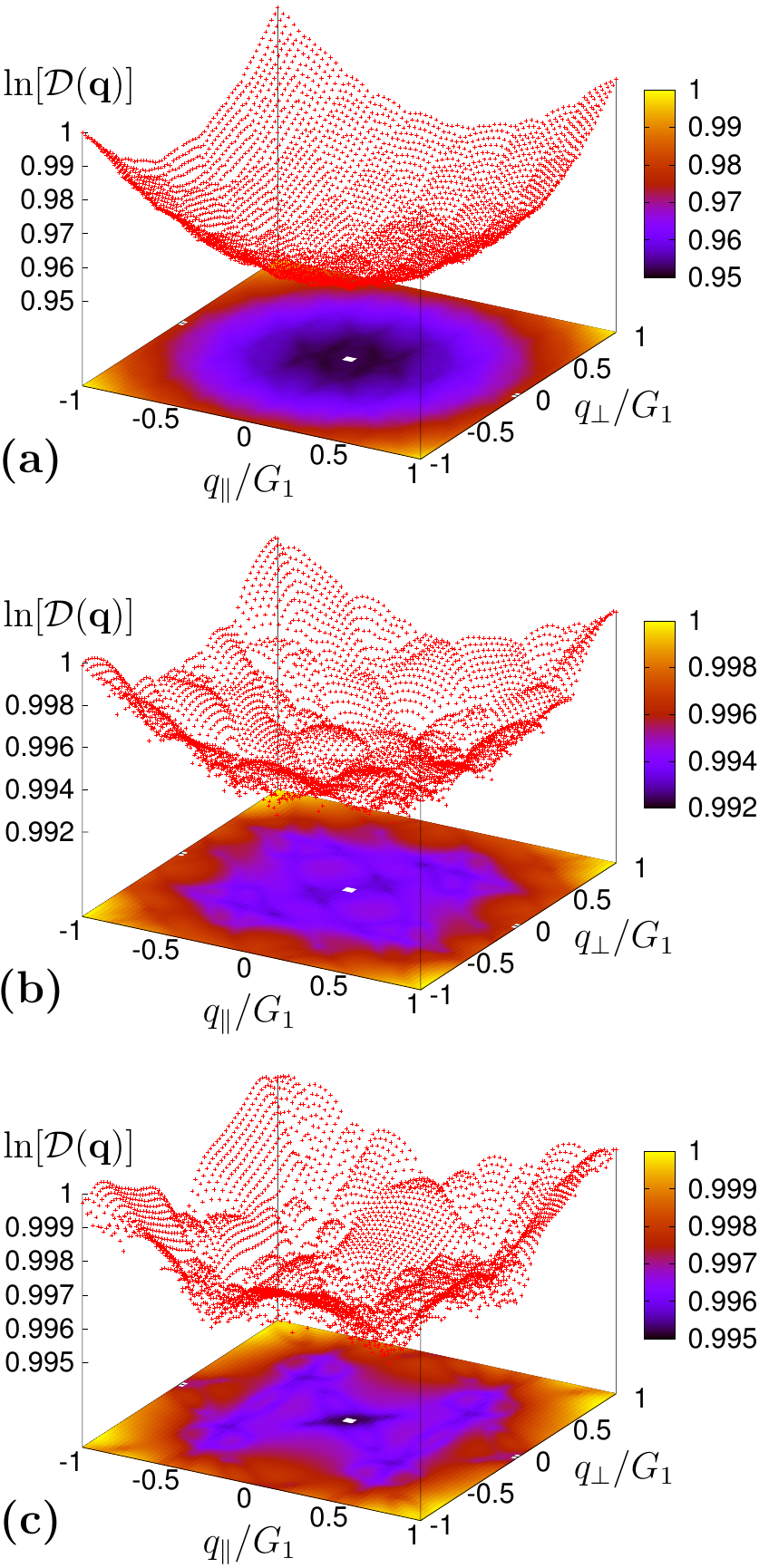}}
\caption{Linear stability of the structures presented in Fig.~\ref{FTstructure} against instabilities of small wave-numbers in the co-moving frame. The analyzed regions in $\mathbf{q}$-space contain the first Brillouin zones of the textures in Fig.~\ref{FTstructure}, respectively. Surface plots show the logarithm of the determinant $\mathcal{D}(\mathbf{q})$ of the stability matrix in the analyzed $\mathbf{q}$-regions. It has been rescaled for technical reasons in each panel, so that the absolute values should not be interpreted. Below the surfaces, the shadow plots represent colored projections of the surfaces into the plane. We define $G_1=\|\mathbf{G}_1\|$, $q_{\|}$ the component of $\mathbf{q}$ parallel to $\mathbf{G}_1$, and $q_{\bot}$ its component in the perpendicular direction, where $\mathbf{G}_1$ is the reciprocal lattice vector of smallest magnitude. A linear instability would be indicated by a vanishing $\mathcal{D}(\mathbf{q})$, however, we observe nonzero values in all probed instances (and no sign changes). We thus conclude that the structures are linearly stable against static instabilities in the co-moving frame.}
\label{logdeterminantco}
\end{figure}

The results corresponding to the three textures in Fig.~\ref{FTstructure} are depicted in Fig.~\ref{logdeterminantco}. 
The determinant naturally vanishes at $\mathbf{q}=\mathbf{0}$ and at $\mathbf{q}=\pm\mathbf{G}_1$, which cannot be shown in the logarithmic plot and leads to the white wholes in the projected shadow plots below each surface plot. 
We do not find any other position in the first Brillouin zone where the determinant vanishes. The small dips visible in the surfaces of $\ln[\mathcal{D}(\mathbf{q})]$ were found to decrease with increasing size of the numerical calculation box that was used to obtain $\psi_{10}$. They were not observed to indicate a linear instability. We thus conclude that the investigated structures are linearly stable against static instabilities in the co-moving frame. 

After excluding static linear instabilities in the co-moving frame, we ask for dynamical linear instabilities. For example, on larger length scales, there might appear domains migrating in different directions. This question adds a further degree of freedom to our analysis, namely the frequency $\omega$. Previously, it was fixed by the relation $\omega=-\mathbf{q}\cdot\mathbf{v_m}$ in the co-moving frame. Now we can in principle choose any value independently of the selected wave vector $\mathbf{q}$. 

Fortunately, for each perturbation $\delta\psi_1(\mathbf{q},\omega)$, the coupled system of equations has the same size as above in the analysis within the co-moving frame. The additional degree of freedom $\omega$ does not lead to further couplings. We can understand this result from Eqs.~(\ref{stabaneq}), (\ref{psi10co}), and (\ref{psi10sqco}). On the one hand, Eq.~(\ref{stabaneq}) implies that the modes $\delta\psi_1$ are coupled via the unperturbed steady-state fields $\psi_{10}(\mathbf{q},\omega)$ and $\psi_{10}^2(\mathbf{q},\omega)$. On the other hand, due to the steady-state form of $\psi_{10}(\mathbf{r},t)$ and $\psi_{10}^2(\mathbf{r},t)$, Eqs.~(\ref{psi10co}) and (\ref{psi10sqco}) contain a rigid relation between $\mathbf{q}$ and $\omega$ in the $\delta$-func\-tions. In this way, each coupling wave vector $\mathbf{q}-\mathbf{q'}$ in Eq.~(\ref{stabaneq}) uniquely fixes one coupling frequency $\omega-\omega'$. 

\begin{figure}
\centerline{\includegraphics[width=8.5cm]{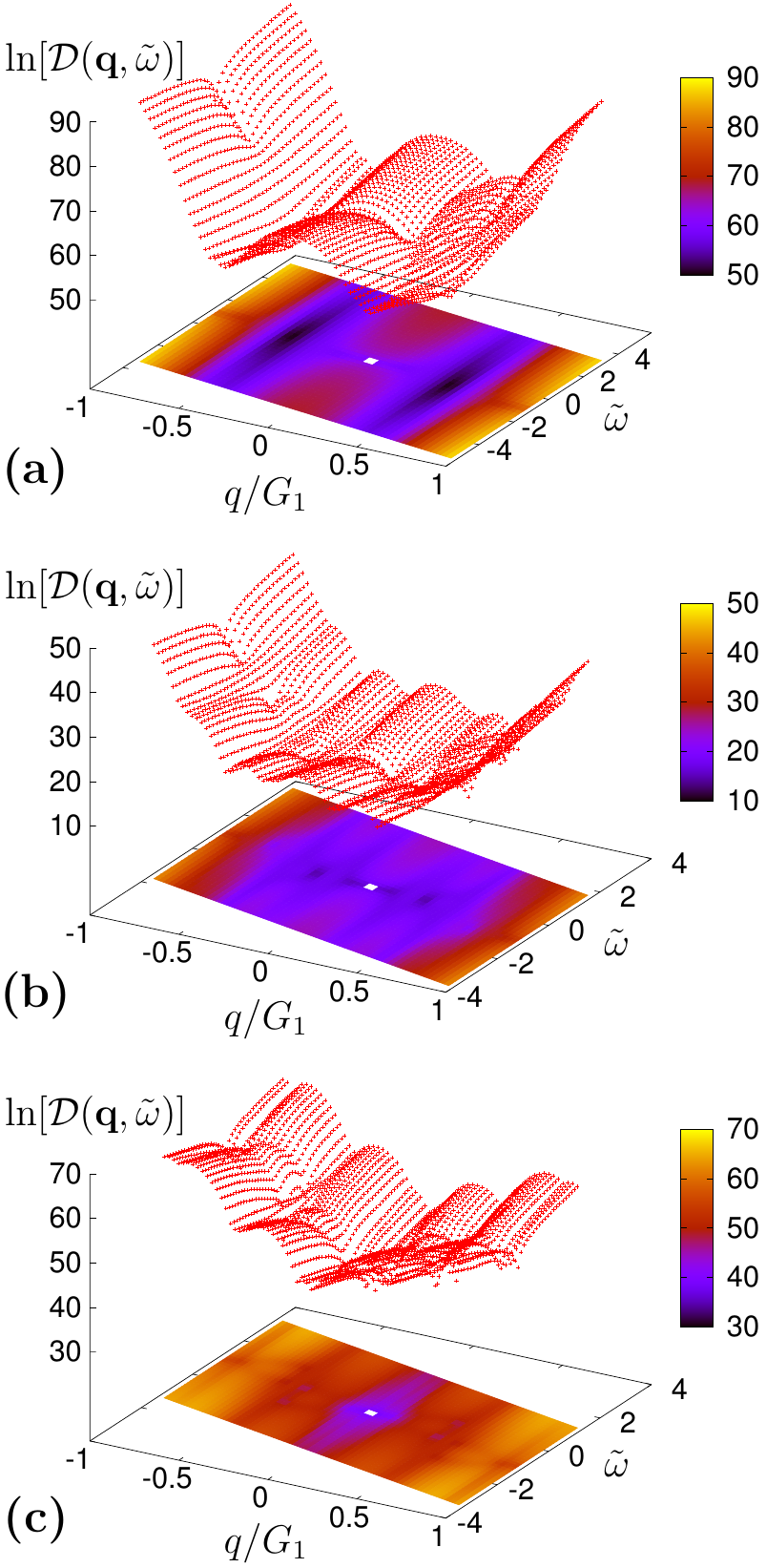}}
\caption{Examples for the linear stability of the structures presented in Fig.~\ref{FTstructure} against linear dynamical instabilities. The panels show stability probes performed in $\mathbf{q}$-space along the direction of the reciprocal lattice vector of smallest magnitude $\mathbf{G}_1$. In addition, the frequency $\omega$ was varied in an interval centered around $-\mathbf{q}\cdot\mathbf{v_m}$ at each wave vector $\mathbf{q}$, with $\mathbf{v_m}$ the collective migration velocity. We define $q=\|\mathbf{q}\|$, $G_1=\|\mathbf{G}_1\|$, and $\tilde{\omega}=\omega/|\mathbf{G}_1\cdot\mathbf{v_m}|$. 
Surface plots show the logarithm of the determinant $\mathcal{D}(\mathbf{q},\tilde{\omega})$ of the stability matrix in the analyzed $\mathbf{q}$-$\tilde{\omega}$ regions. It has been rescaled for technical reasons in each panel, so that the absolute values should not be interpreted. 
Below the surfaces, the shadow plots represent colored projections of the surfaces into the plane. 
Our examples correspond to the directions indicated by the circles in the power spectra of Fig.~\ref{FTstructure} for (a) $v_0=0.5$, (b) $v_0=0.7$, and (c) $v_0=0.9$. 
A linear instability would be indicated by a vanishing $\mathcal{D}(\mathbf{q},\tilde{\omega})$ (or a sign change). Apparently, the structures are linearly stable against linear dynamical instabilities.}
\label{logdeterminantdynamic}
\end{figure}

The complete $\mathbf{q}$-$\omega$ space is too large to be systematically tested. We therefore particularly focused on the representative directions given by the reciprocal wave vectors of each structure, and the directions perpendicular to these. In neither case did we find a vanishing determinant that would indicate a dynamical linear instability. This provides evidence that the investigated traveling active single crystals are also dynamically linearly stable. A few examples of our results are depicted in Fig.~\ref{logdeterminantdynamic}. 

One might argue that there are sources of uncertainty in our analysis. For example, we obtain our initial reciprocal lattice vectors from a numerical calculation on a grid of finite size. However, if this approximation were problematic, it would rather show up as an instability of the unperturbed state. In contrast to that, we here find that the system is linearly stable despite such approximations. We are therefore confident that our analysis reflects the real behavior of the structures, and that the active single crystals are linearly stable against perturbations.

\section{Destabilization via hydrodynamic interactions}

In the previous section, we have seen that the active single crystals are linearly stable even at very large system sizes. The underlying stabilization mechanism is provided by the effective elastic interactions between the density peaks. It is introduced by the phase field crystal free energy functional that acts like a chemical potential for the density peaks in our dynamical equations (\ref{eqpsi1}) and (\ref{eqP}). This picture is further supported by an analysis of related particle simulations \cite{ferrante2013elasticity,ferrante2013collective}. 

On the contrary the question arises, whether in real experimental systems there might be a counteracting mechanism that could destroy this long-ranged order. A natural candidate is given by hydrodynamic interactions. It turns out that hydrodynamic interactions can destabilize the system already at relatively small system sizes. This can be studied by direct numerical simulation of our dynamical equations, so that a complicated analysis as in the previous section is not necessary. 

In the situation that we investigate, the self-propelled particles can still directly exchange momentum with the ground. This requirement is satisfied by particles that directly migrate on a substrate, or approximately by swimmers that propel in a low Reynolds number environment close to a surface with no-slip boundary conditions. However, we assume that the density peaks and lamellae can additionally interact with each other through a surrounding background fluid film. 

Our goal in the following is a simple qualitative estimate of the effect of hydrodynamic interactions. We assume that the particles and the fluid have the same mass density, and that the system is incompressible. In this way the dynamical equation of mass continuity is trivially satisfied and does not need to be considered explicitly in the following. The location of the particles is described by the concentration or density field $\psi_1(\mathbf{r})$. 

We introduce a velocity field $\mathbf{u}(\mathbf{r})$ to parameterize the fluid flow that is induced by the particle motion. 
On the one hand, additional convective terms appear in the dynamical equations (\ref{eqpsi1}) and (\ref{eqP}) that now read 
\begin{eqnarray}
\hspace{-.7cm}\partial_{t}\psi_1 &=& \nabla^2\frac{\delta\mathcal{F}}{\delta\psi_1} - v_0\,\nabla\cdot\mathbf{P} -\mathbf{u}\cdot\nabla\psi_1, 
\label{eqpsi1hydro}\\[.1cm]
\hspace{-.7cm}\partial_{t}\mathbf{P} &=& \nabla^2\frac{\delta\mathcal{F}}{\delta\mathbf{P}} - D_r\frac{\delta\mathcal{F}}{\delta\mathbf{P}} - v_0\nabla\psi_1 -\mathbf{u}\cdot\nabla\mathbf{P} +\mathbf{\Omega}\cdot\mathbf{P}.
\label{eqPhydro}
\end{eqnarray}
Here the tensor $\mathbf{\Omega}$ with components $\Omega_{ij}=(\nabla_iu_j-\nabla_ju_i)/2$ describes rotations due to convection. We do not consider flow alignment of the polarization field in this qualitative picture \cite{degennes2003physics,aditi2002hydrodynamic}. 

On the other hand, the additional dynamical equation for the velocity field $\mathbf{u}(\mathbf{r})$ in the rescaled form reads
\begin{equation}
\partial_t\mathbf{u} = -\mathbf{u}\cdot\nabla\mathbf{u} + \mathbf{F} + \nabla\cdot\nu\{\psi_1\}\nabla\mathbf{u} -\alpha\mathbf{u}.    
\label{equ}
\end{equation}
It is supplemented by the incompressibility condition $\nabla\cdot\mathbf{u}=0$, so that an explicit pressure term is not necessary. We can neglect the convective term on the right-hand side in the low Reynolds number regime. 

When the self-propelled particles migrate on the substrate, they push the surrounding fluid and set it into motion. This effect is included by the force term $\mathbf{F}$ that we introduce as 
\begin{equation}
\mathbf{F} = \gamma \sum_{i=1}^{N_p} (\mathbf{v}_i-\mathbf{u})\,\delta(\mathbf{r}-\mathbf{R}_i)
\end{equation}
with a sufficiently large coefficient $\gamma>0$. Again $N_p$ is the number of density peaks in the sample. $\mathbf{R}_i$ and $\mathbf{v}_i$ denote, respectively, the current positions and velocities of the density peaks for $i=1,\dots,N_p$. In practice we track the positions and motion of the centers of the density peaks as described above. At each peak center $\mathbf{R}_i$, the fluid flow velocity $\mathbf{u}$ is then adjusted to the peak velocity $\mathbf{v}_i$. To mimic the presence of the particles in the fluid, the viscosity $\nu$ varies with the density $\psi_1$. We set $\nu=10$ at positions of lowest density $\psi_1$ and let it linearly increase by a factor of $2$--$3$ when moving to the regions of high density $\psi_1$. Due to the smooth density profiles in our numerical samples the effect of the contribution $(\nabla\nu\{\psi_1\})\cdot\nabla\mathbf{u}$ is mostly negligible. Overall, the volume elements of high density $\psi_1$, which represent the presence of the self-propelled particles, are considered as part of the fluid flow. This approach is similar to the ``fluid particle dynamics'' introduced by Tanaka and Araki \cite{tanaka2000simulation}. 

The last term $-\alpha\mathbf{u}$ in Eq.~(\ref{equ}) represents a simple way to model the friction of the thin fluid film with its environment, with $\alpha>0$ the friction constant. This term was used previously to investigate the complex flow behavior in quasi-two-dimensional systems of fluid films \cite{rivera2000external}. Two limits for the magnitude of $\alpha$ are obvious. On the one hand, if $\alpha$ is small, the friction between the fluid film and its surfaces is low and the migrating particles can easily set the surrounding fluid into motion. On the other hand, if $\alpha$ is very large, we see from Eq.~(\ref{equ}) that the fluid remains practically at rest with $\mathbf{u}(\mathbf{r})\approx\mathbf{0}$. Then the convective terms in Eqs.~(\ref{eqpsi1hydro}) and (\ref{eqPhydro}) are negligible and we reobtain the previous Eqs.~(\ref{eqpsi1}) and (\ref{eqP}). Thus hydrodynamic interactions are not important in the regime of large $\alpha$. 

\begin{figure*}
\centerline{\includegraphics[width=17.8cm]{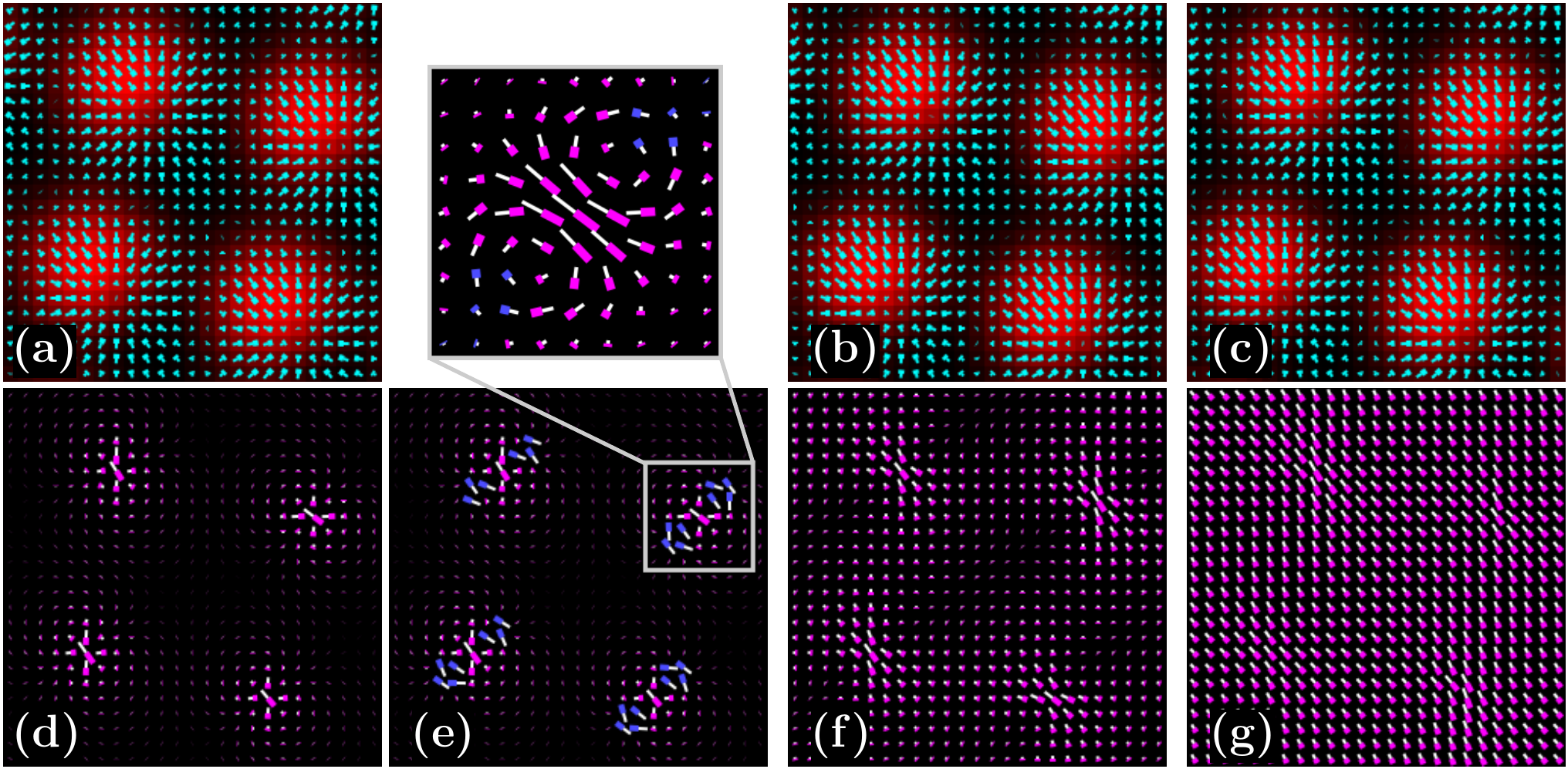}}
\caption{Hydrodynamic interactions for the two-dimensional self-propelled motion on or close to a surface in a surrounding fluid film at $(\bar{\psi},\varepsilon,C_1,C_4,v_0)=(-0.4,-0.98,0.2,0,0.7)$. The top panels (a)--(c) show the density field $\psi_1$ and the polarization field $\mathbf{P}$ in the same way as in Fig.~\ref{snapshots}. We illustrate the corresponding fluid flow fields $\mathbf{u}$ in the bottom panels (d), (f), and (g) by needles pointing from the thick to the thin white ends. The friction parameter $\alpha$ of the fluid film decreases from the left to the right column: (a, d, e) $\alpha=300$, (b, f) $\alpha=10$, (c, g) $\alpha=1.5$. Hydrodynamic interactions are weak at high fluid friction, where the flow remains mostly localized around the density peaks (a,d). On the contrary, hydrodynamic interactions are strong at low fluid friction, where the whole fluid can be set into motion nearly homogeneously and the fluid flow extends between the density peaks (c,g). Panel (e) highlights the ``backflow'' regions occurring in panel (d) around the density peaks by rescaling the magnitude of the depicted flow vectors. A ``zoom'' into the region around one single density peak is included for illustration. Only a fraction of the numerical calculation box is shown in each case.}
\label{fluidflow}
\end{figure*}

We illustrate the flow behavior at different values of the friction parameter $\alpha$ in Fig.~\ref{fluidflow}. 
For high $\alpha$, corresponding to strong fluid friction, most parts of the fluid film cannot be set into motion as illustrated in panels (a) and (d). The fluid flow remains very localized around the density peaks and does not extend between different peaks. Consequently the hydrodynamic interactions between the density peaks are low. Around the density peaks some ``backflow'' of the fluid oppositely to the peak migration directions occurs. This is stressed in a rescaled version of panel (d), included as panel (e); a ``zoom'' into the region around one single density peak is provided for illustration. With decreasing fluid friction $\alpha$, the ``backflow'' vanishes and the fluid flow more and more extends between the density peaks, as depicted in panels (b) and (f). At very low friction parameters $\alpha$, the whole fluid can be set into motion nearly homogeneously, as shown in panels (c) and (g). Since the fluid flow here extends between the density peaks, hydrodynamic interactions are strong in this case. 

\begin{figure}
\centerline{\includegraphics[width=8.5cm]{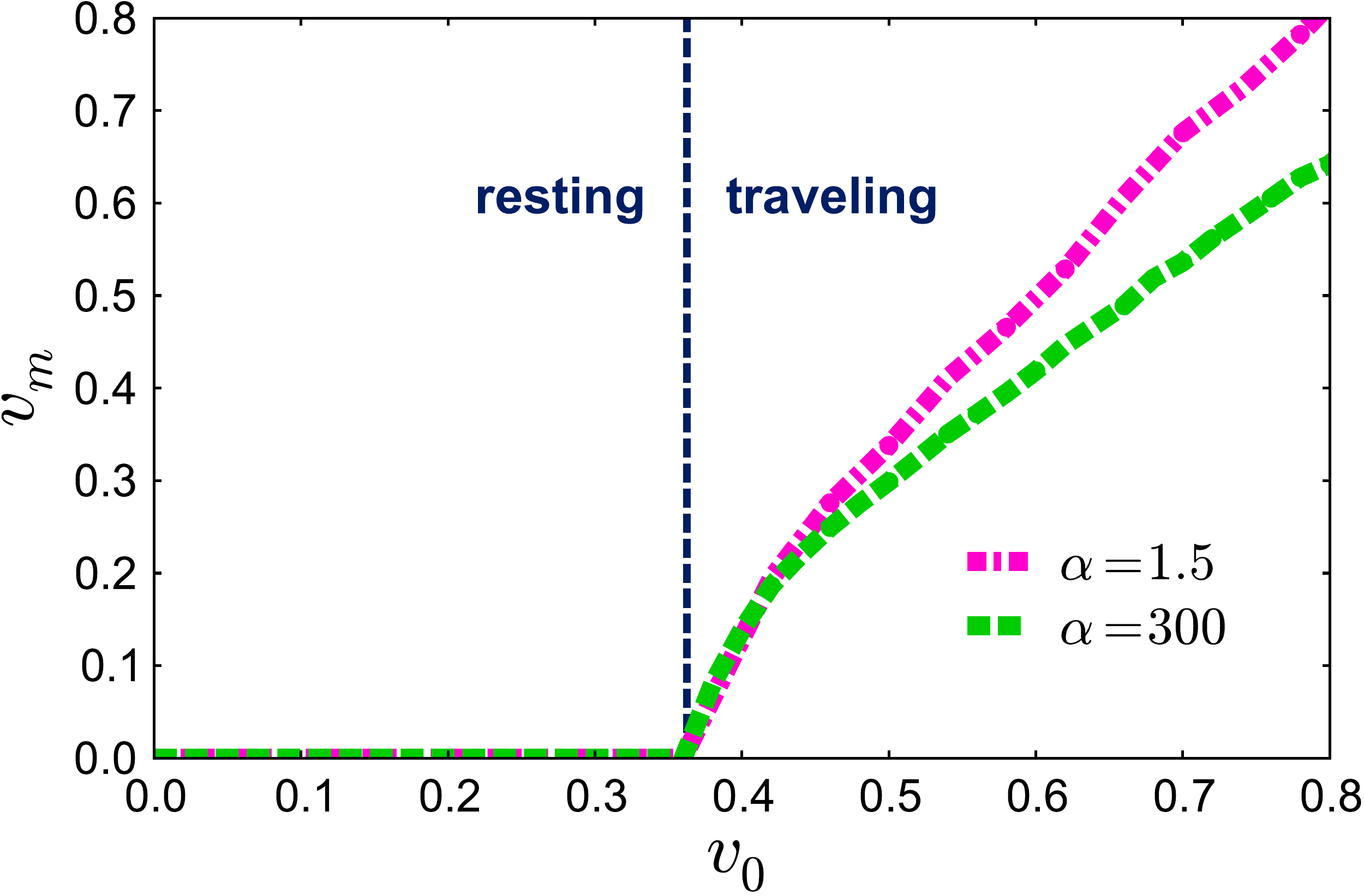}}
\caption{Collective migration speed $v_m$ of the single crystals as a function of the active drive $v_0$ of the individual particles in the presence of hydrodynamic interactions at $(\bar{\psi},\varepsilon,C_1,C_4)=(-0.4,-0.98,0.2,0)$. For low hydrodynamic interactions, corresponding to a large fluid friction parameter $\alpha$, the collective migration speed is lower. For strong hydrodynamic interactions, corresponding to a low fluid friction parameter $\alpha$, the collective migration speed is enhanced. In the latter case the particles can push each other by setting the fluid between them into motion. The critical value of the self-propulsion velocity $v_{0,c}$ remains unchanged by the presence of the hydrodynamic interactions. 
}
\label{alphadep}
\end{figure}

As for Fig.~\ref{vmvsv0}, we measured again the collective migration speed $v_m$ in the form of the sample-averaged peak velocity magnitude, $v_m=\sum_{i=1}^{N_p} \|\mathbf{v}_i\|/{N_p}$, with $N_p$ the number of the density peaks and $\mathbf{v}_i$ the velocity of the single density peaks, $i=1,\dots,N_p$. A result of $v_m$ as a function of the active drive $v_0$ of the individual self-propelled particles is plotted in Fig.~\ref{alphadep} for two different 
values of the fluid friction parameter $\alpha$. 
In this case $C_1>0$, i.e.\ there is no alignment mechanism that would lead to a spontaneous orientational ordering of the self-propulsion directions. 

As a first result, we found that, within our numerical resolution, the critical self-propulsion velocity $v_{0,c}$, at which collective motion of the active crystal sets in, remains unchanged by the hydrodynamic interactions. Second, the hydrodynamic interactions speed up the collective motion of the active single crystal. We can see this from the two data curves plotted in Fig.~\ref{alphadep}. At high values of $\alpha$ the fluid friction is large, and the self-propelling particles cannot set the surrounding fluid into motion. However, at low fluid friction for low values of $\alpha$, the particles can set the whole surrounding fluid film into motion, as was also demonstrated in Fig.~\ref{fluidflow}. In this way, one density peak can push the preceding peak via the fluid between them. The density peaks support each other in migration via hydrodynamic interactions. At fixed active drive $v_0$ of the individual self-propelled particles, the collective migration speed $v_m$ increases with decreasing fluid friction $\alpha$ and increasing hydrodynamic interactions. 

\begin{figure}
\centerline{\includegraphics[width=8.5cm]{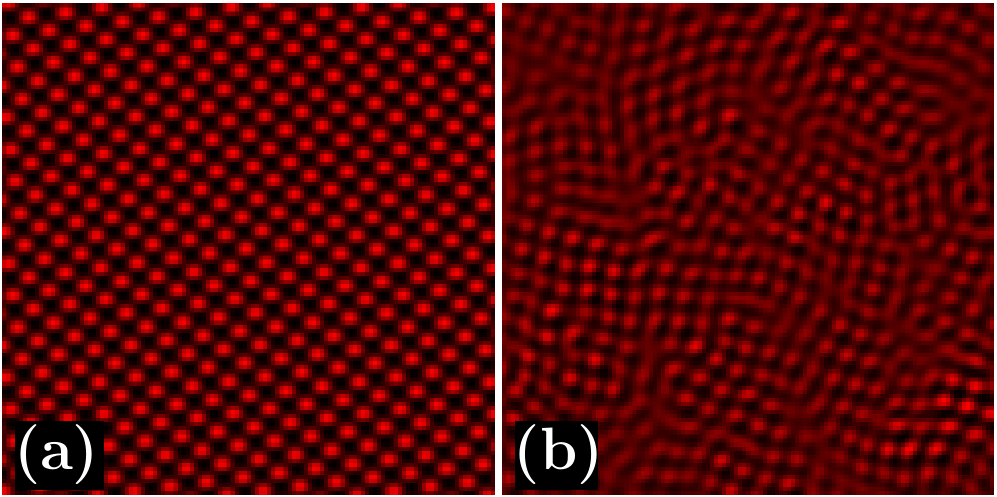}}
\caption{Hydrodynamic interactions can destabilize a collectively migrating active single crystal at $(\bar{\psi},\varepsilon,C_1,C_4,v_0)=(-0.4,-0.98,0.2,0,1)$. (a) Negligible hydrodynamic interactions at strong fluid friction ($\alpha=300$) allow the emergence of a perfect traveling active single crystal. (b) Strong hydrodynamic interactions at low fluid friction ($\alpha=1.5$) destabilize this texture.}
\label{destabilizedcrystal}
\end{figure}

However, the hydrodynamic interactions disturb the order of the active single crystals. This is illustrated in Fig.~\ref{destabilizedcrystal}. 
The figure includes two identical systems that only differ by the fluid friction parameter $\alpha$. At high $\alpha$ and therefore little hydrodynamic interaction (a), the sample shows a perfect active single crystal that migrates collectively into one direction. On the contrary, the crystal is broken up into different domains at low $\alpha$ corresponding to strong hydrodynamic interactions (b). Also a partial transition to lamellar textures is induced by the hydrodynamic interactions. In the case without hydrodynamic interactions this transition would typically occur only at higher magnitudes of the active drive $v_0$ \cite{menzel2013traveling}. 

In summary, we can say that -- at least within the qualitative picture presented -- hydrodynamic interactions have a destabilizing effect on the active crystalline structures considered here.

\section{Conclusions}

In this paper we focused on the stability of traveling active single crystals. An active phase field crystal model was introduced for this purpose. We outlined its derivation via dynamical density functional theory from a microscopic consideration of self-propelled particles that feature an active drive. Besides resting hexagonal crystals, traveling hexagonal, rhombic, and quadratic crystals, as well as traveling lamellar textures were obtained from this description. Also the inverted structures like resting and traveling honeycomb textures, as well as inverted traveling rhombic and quadratic lattices were observed. A linear stability analysis did not indicate a linear instability of the collectively migrating single crystals. In particular, long-wave-length instabilities beyond the size of the numerical calculation grids did not show up, indicating linear stability also for large system sizes. We found, however, that hydrodynamic interactions can destabilize the single-crystalline textures. 

The linear stability analysis was significantly complicated by the spatial modulation of the unperturbed steady state. As a consequence of this spatial modulation, our equations became nonlocal in Fourier space. 
We could reduce the problem to manageable size by exploiting the periodicity of the unperturbed state and its steady-state migration. 
This made the linear stability analysis possible. 
The problem of a spatial modulation of the unperturbed state occurred already in a previous equilibrium stability study of micro-phase separated diblock copolymer melts \cite{laradji1997stability}. There, however, the equilibrium conditions allowed a different strategy to perform the analysis. 

Our results on the stability of the traveling crystals are important when we think of the design of new active materials. It is expected that, for example, features of traveling crystals like phononic properties are different from their passive counterparts. We have seen that the structure of the migrating crystals, i.e.\ their hexagonal, rhombic, or quadratic form, can be tuned by the activity of the constituent self-propelled particles. It will be interesting to test whether other features of active crystals, such as their phononic behavior, can be similarly tuned by the properties of the individual active components. We shall investigate this interesting question in the future.

\begin{acknowledgments}
The authors thank the Deutsche Forschungsgemeinschaft for support of this work through the German-Japanese project LO 418/15. 
\end{acknowledgments}

\appendix*

\section{One-dimensional analytical approach to the onset of collective motion}

In this appendix we briefly analytically address the onset of collective migration. Our goal is to derive an expression for the collective migration speed $v_m$. This is a challenging task. We confine ourselves to one spatial dimension that we denote as the $z$-direction. Furthermore, we consider the case of $C_1>0$ and $C_4=0$. Still our results remain restricted to the very close vicinity of the threshold for the onset of collective motion. 

Under these circumstances, and using the fact that the structures collectively migrate in a steady-state solution with speed $v_m$, see Eq.~(\ref{steadystate}), we can rewrite Eqs.~(\ref{eqpsi1}) and (\ref{eqP}) as 
\begin{eqnarray}
-v_m\partial_z\psi_1 &=& \partial_z^2\Big\{\!\left[\varepsilon+(1+\partial_z^2)^2
+3\bar{\psi}^2\right]\psi_1 
\label{apppsi1}
\nonumber\\
& & {}\qquad+ 3\bar{\psi}\psi_1^2+\psi_1^3\,\Big\}  - v_0\,\partial_z{P}, \\[.1cm]
-v_m\partial_z{P} &=& C_1\!\left( \partial_z^2-D_r \right)\!P - v_0\partial_z\psi_1.
\label{appP}
\end{eqnarray}

Introducing the abbreviations $\bar{v}_0=v_0/C_1$ and $\bar{v}_m=v_m/C_1$, we can write the formal solution of Eq.~(\ref{appP}) as 
\begin{eqnarray}
P & = & -\bar{v}_0\left(-\partial_z^2+D_r-\bar{v}_m\partial_z\right)^{-1}\partial_z\psi_1
\nonumber\\
&=& {} -\bar{v}_0\left[ \Gamma+\bar{v}_m\Gamma^2 \partial_z
  +\bar{v}_m^2\Gamma^3\partial_z^2+\bar{v}_m^3\Gamma^4\partial_z^3+...\right]
  \partial_z\psi_1 
\nonumber\\
&=& {} -\bar{v}_0\partial_z\left[ \Gamma+\bar{v}_m\Gamma^2 \partial_z
  +\bar{v}_m^2\Gamma^3\partial_z^2+\bar{v}_m^3\Gamma^4\partial_z^3+...\right]\psi_1, 
\nonumber\\&&\label{solP}
\end{eqnarray}
where we have defined $\Gamma$ via
\begin{equation}
\left(-\partial_z^2+D_r\right)\Gamma = \delta(z-z')
\end{equation}
with the abbreviation
\begin{equation}
\Gamma X(z) = \int dz'\,\Gamma(z-z')X(z') \label{abbrGamma}
\end{equation}
when the operator $\Gamma$ is applied to an arbitrary function $X(z)$. We have further used the relation
\begin{eqnarray}
\int dz' \,\Gamma(z-z')\partial_{z'}\psi(z') 
&=& {}-\int dz'\, \left[\partial_{z'}\Gamma(z-z')\right]\psi(z') \nonumber\\
&=& \partial_z \int dz'\,\Gamma(z-z')\psi(z')\nonumber\\
&&
\end{eqnarray}
in the last step of Eq.~(\ref{solP}). 

Eq.~(\ref{solP}) is now inserted into Eq.~(\ref{apppsi1}) to obtain a closed equation for $\psi_1$: 
\begin{eqnarray}
-v_m\partial_z\psi_1 &=& \partial_z^2\Big\{\!\left[\varepsilon+(1+\partial_z^2)^2
+3\bar{\psi}^2\right]\psi_1 
\nonumber\\
& & {}\qquad+ 3\bar{\psi}\psi_1^2+\psi_1^3\,\Big\}   
\nonumber\\
& & {} +v_0\bar{v}_0\partial_z^2\big[ \Gamma+\bar{v}_m\Gamma^2 \partial_z
\nonumber\\
& & {} \qquad+\bar{v}_m^2\Gamma^3\partial_z^2
  +\bar{v}_m^3\Gamma^4\partial_z^3+...\big]\psi_1. 
\nonumber\\
\end{eqnarray}
We rewrite this expression in the form
\begin{eqnarray}
v_mG\partial_z\psi_1 &=& \Big\{\!\left[\varepsilon+(1+\partial_z^2)^2
+3\bar{\psi}^2\right]\psi_1 
\nonumber\\
& & {}\qquad+ 3\bar{\psi}\psi_1^2+\psi_1^3\,\Big\}   
\nonumber\\
& & {} +v_0\bar{v}_0\big[ \Gamma+\bar{v}_m\Gamma^2 \partial_z
\nonumber\\
& & {} \qquad+\bar{v}_m^2\Gamma^3\partial_z^2
  +\bar{v}_m^3\Gamma^4\partial_z^3+...\big]\psi_1, 
\nonumber\\
\label{Gintroduced}
\end{eqnarray}
where we have defined
\begin{equation}
{}-\partial_z^2G = \delta(z-z')
\end{equation}
with an abbreviation analogous to Eq.~(\ref{abbrGamma}). 

Next, we multiply Eq.~(\ref{Gintroduced}) by $\partial_z\psi_1$ and carry out the integral over $z$:
\begin{eqnarray}
\lefteqn{v_m\int dz\,(\partial_z\psi_1)G\partial_z\psi_1=}\nonumber\\
&& \qquad\qquad\int dz\, (\partial_z\psi_1)\left(2\partial_z^2+\partial_z^4\right)\psi_1
\nonumber\\
&& \qquad\qquad{}+ v_0\bar{v}_0\int dz\, (\partial_z\psi_1){}\big[ 
  \Gamma+\bar{v}_m\Gamma^2 \partial_z
\nonumber\\
&& {} \qquad\qquad\qquad\qquad+\bar{v}_m^2\Gamma^3\partial_z^2
  +\bar{v}_m^3\Gamma^4\partial_z^3+...\big]\psi_1.\qquad
\nonumber\\
&& \label{inteq}
\end{eqnarray}
Here the integrals are taken over the whole domain of size $L$. The periodic boundary conditions at the domain boundaries imply 
\begin{equation}
\int dz\,(\partial_z\psi_1)W(\psi_1) = \int dz\,\partial_z\int^{\psi_1(z)}d\xi\,W(\xi)=0
\end{equation}
for any integrable functional $W$. 

Our last step consists of expanding the operator $\Gamma$ on the right-hand side of Eq.~(\ref{inteq}) into powers of $\partial_z^2$. On the one hand, we have
\begin{eqnarray}
\lefteqn{\int dz\,(\partial_z^{2n+1}\psi_1)(\partial_z^{2m}\psi_1)=}\nonumber\\
&& \qquad\frac{(-1)^{n+m+1}}{2}\int dz\,\partial_z(\partial_z^{n+m}\psi_1)^2 =0,
\end{eqnarray}
and on the other hand,
\begin{eqnarray}
\lefteqn{\int dz\,(\partial_z^{2n+1}\psi_1)(\partial_z^{2m+1}\psi_1)=}\nonumber\\
&& \qquad{(-1)^{n+m}}\int dz\,(\partial_z^{n+m+1}\psi_1)^2 \neq0,
\end{eqnarray}
with integers $n\geq0$ and $m\geq0$.  
Using these relations, we finally obtain from Eq.~(\ref{inteq}) an expression for the collective migration speed at the onset of collective motion: 
\begin{equation}
(B_0-B_1)v_m+B_3v_m^3=0, \label{eqvm}
\end{equation}
where
\begin{eqnarray}
B_0 &=& \int dz\,\int dz'\,\left[\partial_z\psi_1(z)\right]G(z-z')\partial_{z'}\psi_1(z'),
\nonumber\\
&&\label{B0}
\\
B_1 &=& \frac{v_0^2}{C_1^2}\int dz\,\int dz'\,\int dz''\,\left[\partial_z\psi_1(z)\right]
\nonumber\\
  && {}\qquad\times\Gamma(z-z')\Gamma(z'-z'')\partial_{z''}\psi_1(z''),
\nonumber\\
&&\label{B1}
\\
B_3 &=& \frac{v_0^2}{C_1^4}\int dz\,\int dz'\,\int dz''\,\int dz'''\,\int dz''''\,
\nonumber\\
  && {}\qquad\times\left[\partial_z^2\psi_1(z)\right]\Gamma(z-z')\Gamma(z'-z'')
\nonumber\\
  && {}\qquad\times\Gamma(z''-z''')\Gamma(z'''-z'''')\partial_{z''''}^2\psi_1(z'''').
\nonumber\\
&&\label{B3}
\end{eqnarray}
From the symmetry of the integrands in these expressions, and since $G$ and $\Gamma$ are positive operators, we expect all coefficients $B_0$, $B_1$, and $B_3$ to be positive. This implies a supercritical bifurcation of the collective migration speed at the onset of collective motion, which is in line with our simulations in two spatial dimensions. 
However, since $\psi_1(z)$ generally depends on $v_m$, the coefficient $B_3$ might be renormalized from the $v_m$-dependence of $B_0$ and $B_1$. 

Directly at the threshold for the onset of collective motion, we may assume a harmonic functional dependence of the form $\psi_1(z)\sim\sin(q_0z)$. When we calculate the coefficients $B_0$, $B_1$, and $B_3$ from Eqs.~(\ref{B0})--(\ref{B3}), we obtain from Eq.~(\ref{eqvm}) for the collective migration speed directly above onset: 
\begin{equation}
|v_m|=\left[ \frac{C_1^2}{q_0^2}\left(D_r+q_0^2\right)^2 - \frac{C_1^4}{v_0^2q_0^4}\left(D_r+q_0^2\right)^4
\right]^{\frac{1}{2}}. 
\end{equation}
However, higher harmonics in the expression for $\psi_1(z)$ and its $v_m$-dependence will affect this result.

\end{document}